%% file: main.tex
\documentclass[conference,9pt]{IEEEtran}
\IEEEoverridecommandlockouts
\usepackage{cite}
\usepackage{amsmath,amssymb,amsfonts}
\usepackage{algorithmic}
\usepackage{graphicx}
\usepackage{textcomp}
\usepackage{xcolor}
\usepackage{caption}
\usepackage{listings}
\usepackage{pxfonts}
\usepackage{subcaption}
\usepackage{braket}
\usepackage[compact]{titlesec}

\usepackage{xcolor}
\definecolor{commentgreen}{RGB}{2,112,10}
\definecolor{eminence}{RGB}{108,48,130}
\definecolor{weborange}{RGB}{255,165,0}
\definecolor{frenchplum}{RGB}{129,20,83}

\usepackage[
  bookmarks=true,
  bookmarksnumbered=true,
  draft=false,
  pagebackref=false,
  colorlinks=true,
  linkcolor=black,
  citecolor=black,
  filecolor=black,
  urlcolor=black,
  pdfpagelabels=false,
]{hyperref}
\usepackage{balance}

\usepackage{tikz}
\newcommand*\circled[1]{\tikz[baseline=(char.base)]{
            \node[shape=circle,draw,inner sep=0.5pt] (char) {#1};}}

\def\BibTeX{{\rm B\kern-.05em{\sc i\kern-.025em b}\kern-.08em
    T\kern-.1667em\lower.7ex\hbox{E}\kern-.125emX}}

\begin{document}
\pagestyle{plain}
\title{Quantum Computing in the Cloud: \\ Analyzing job and machine characteristics\\
}

\author{\IEEEauthorblockN{Gokul Subramanian Ravi}
\IEEEauthorblockA{University of Chicago\\
gravi@uchicago.edu}
\and
\IEEEauthorblockN{Kaitlin N. Smith}
\IEEEauthorblockA{University of Chicago\\
kns@uchicago.edu}
\and
\IEEEauthorblockN{Pranav Gokhale}
\IEEEauthorblockA{Super.tech\\
pranav@super.tech}
\and
\IEEEauthorblockN{Frederic T. Chong}
\IEEEauthorblockA{University of Chicago\\
chong@cs.uchicago.edu}

}

\maketitle
\pagenumbering{gobble}

\input{0_abstract}

\input{1_introduction}
\input{2_background}

\input{3_obsA}

\input{4_obsB}

\input{5_obsC}

\input{6_obsD}

\input{9_backend}

\section*{Acknowledgement}
This work is funded in part by EPiQC, an NSF Expedition in Computing, under grants CCF-1730082/1730449; in part by STAQ under grant NSF Phy-1818914; in part by NSF Grant No. 2110860; in part by the US Department of Energy Office  of Advanced Scientific Computing Research, Accelerated Research for Quantum Computing Program; and in part by  NSF OMA-2016136 and in part based upon work supported by the U.S. Department of Energy, Office of Science, National Quantum Information Science Research Centers.  
GSR is supported as a Computing Innovation Fellow at the University of Chicago. This material is based upon work supported by the National Science Foundation under Grant \# 2030859 to the Computing Research Association for the CIFellows Project.
KNS is supported by IBM as a Postdoctoral Scholar at the University of Chicago and the Chicago Quantum Exchange.
FTC is Chief Scientist at Super.tech and an advisor to Quantum Circuits, Inc.

\bibliographystyle{IEEEtranS}
\balance
\bibliography{refs}

\end{document}

%% file: 0_abstract.tex
\begin{abstract}
As the popularity of quantum computing continues to grow, quantum machine access over the cloud is critical to both academic and industry researchers across the globe.
And as cloud quantum computing demands increase exponentially, the analysis of resource consumption and execution characteristics are key to efficient management of jobs and resources at both the vendor-end as well as the client-end.
While the analysis of resource consumption and management are popular in the classical HPC domain, it is severely lacking for more nascent technology like quantum computing.

This paper is a first-of-its-kind academic study, analyzing various trends in  job execution and  resources consumption / utilization on quantum cloud systems.
We focus on IBM Quantum systems and analyze characteristics over a two year period, encompassing over 6000 jobs which contain over 600,000 quantum circuit executions and correspond to almost 10 billion "shots" or trials over 20+ quantum machines.
Specifically, we analyze trends focused on, but not limited to, execution times on quantum machines, queuing/waiting times in the cloud, circuit compilation times, machine utilization, as well as the impact of job and machine characteristics on all of these trends.
Our analysis identifies several similarities and differences with classical HPC cloud systems.
Based on our insights, we make recommendations and contributions to improve the management of resources and jobs on future quantum cloud systems.
\end{abstract}

%% file: 1_introduction.tex
\section{Introduction}
\label{Introduction}

Quantum computing is a revolutionary computational model that leverages quantum mechanical phenomena for solving intractable problems. 
Quantum computers (QCs) evaluate quantum circuits or programs in a manner similar to a classical computer, but quantum information's ability to leverage superposition, interference, and entanglement gives QCs significant advantages in cryptography~\cite{shor1999polynomial}, chemistry~\cite{kandala2017hardware}, optimization~\cite{moll2018quantum}, and machine learning~\cite{biamonte2017quantum}.

With development of today's Noisy Intermediate-Scale Quantum (NISQ)  devices, cloud-based Quantum Information Processing platforms with nearly 100 qubits are currently accessible to the public.
Further, recent quantum hardware roadmaps, such as IBMs~\cite{IBM-HW}, have announced that devices with as many as 1000 qubits will be available by 2023. 
It also has been recently demonstrated by the Quantum Supremacy experiment on the Sycamore quantum processor, a 54-qubit quantum computing device manufactured by Google, that quantum computers can outperform current classical supercomputers in certain computational tasks~\cite{arute2019quantum}. 
These developments suggest that the future of quantum computing is promising.

\begin{figure}[t]
\fbox{
\includegraphics[width=\columnwidth,trim={0cm 0cm 0cm 0cm},clip]{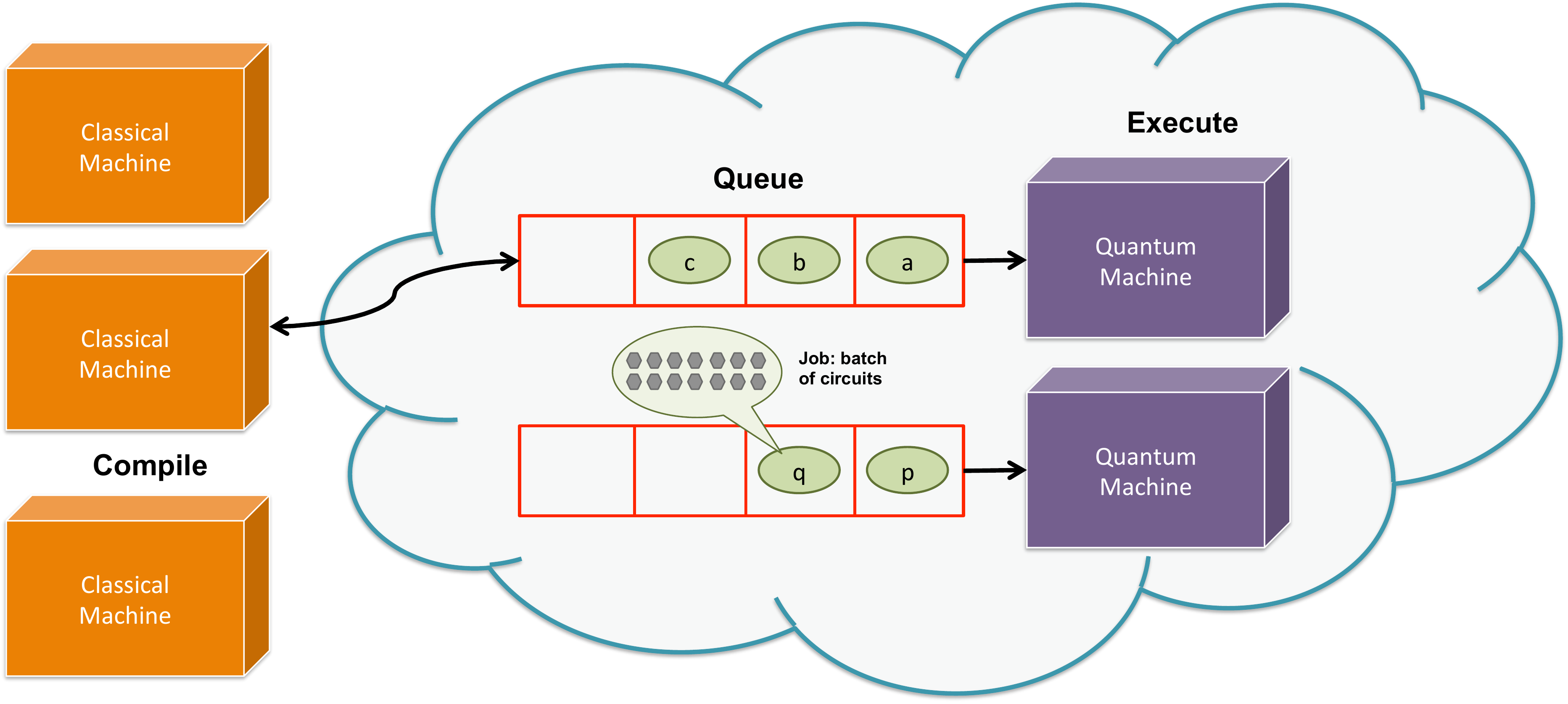}
}
\centering
\caption{Quantum in the cloud. Clients launch quantum programs from their classical computers onto the vendor's quantum cloud wherein the jobs are queued until execution.} 
\label{Fig:SC21_Overview}
\end{figure}

While the future looks promising, quantum computing is still at a nascent stage and quantum computers are a rare and expensive resource.
Thus, quantum machines and corresponding software stacks are primarily accessed by researchers in academia and industry world wide via the cloud.
Current cloud vendors with their own quantum hardware include industry giants like IBM, Google, Microsoft and Honeywell, as well as startups such as Xanadu, Rigetti, IonQ and D-Wave (Note: D-Wave's quantum annealer is different from a traditional quantum computer).
Further, Amazon Braket (AWS) and Microsoft Azure Quantum provide quantum computing as a service via multiple other quantum hardware vendors.
It is expected that quantum computing as a cloud service will grow considerably over the next decade and will continue to be the main access to quantum machines for research across the globe.
Fig.\ref{Fig:SC21_Overview} provides an overview of how clients interact with cloud quantum machines - more details are discussed in Section \ref{Background}.

In the current quantum era, there is growing scarcity of quantum resources even in the cloud, as the demand consistently increases. 
Quantum machines  available in the cloud are very limited in number~\cite{IBMQE,AWS,Azure}, and the number of users and the number of ``jobs” submitted to these machines are drastically growing every day~\cite{IBM-users} across multiple vendors. 
With the increasing popularity of quantum computing in both industry and academic research, it is expected that these contention trends will continue to worsen over the next decade or more - at the very least until the cost of building large and reliable quantum computers becomes more easily surmountable.
As an example, a first-order impact of quantum machine scarcity are the long queuing times~\cite{das:2019,kong2021origin} experienced while accessing cloud machines. 
As discussed in later sections, we observe that there can be 10s-1000s of quantum jobs queued up on quantum machines at any given time. 
This results in queuing times of many hours and sometimes even days.
Such accessibility constraints in using these machines can severely handicap several research endeavors in terms of: a) the scope of the quantum problems that can be effectively targeted on these machines, and b) timely access to the machines irrespective of the quantum problem.

Thus, as quantum demand continuous to grow, it is imperative to efficiently manage quantum resources.
Similar to classical HPC, vendors should try to allocate machine resources as efficiently as possible so as to improve system throughput, while clients should try to make efficient use of job deployment strategies to maximize their allocated time and resources.
Unlike classical HPC though, on the one hand, quantum machines are significantly impacted by machine fidelity constraints (such as static qubit connectivity within the machines and dynamic qubit error rates) meaning that machine utilization cannot be naively maximized, and on the other hand, in the near future quantum jobs / circuits are expected to be on the lower end of the complexity spectrum, meaning that their execution characteristics can be more easily predictable.

To obtain and understand such insights, it is critical to understand the characteristics of the executing quantum jobs as well as those of the machines in the cloud.
In the HPC world, large scale system characterization are important in building and investing in the next generation of computing systems.
We expect that quantum cloud systems will follow suit.

In this study, we analyze quantum executions on more than 20 IBM Quantum Computers~\cite{IBMQE}, over a 2 year period up to April 2021.
Our study includes over 6000 jobs run on these quantum machines, which encompass over 600,000 quantum circuits.
With each circuit being run for multiple trials / shots on the quantum machines (for higher confidence), our study includes results from as many as 10 billion machine executions over this period.
Note that our data corresponds to quantum experiments run in an academic research setting.
But our insights and recommendations are widely applicable to general quantum cloud systems.

\textbf{Novel insights and recommendations in this paper include:}

    \circled{1}\ \emph{As quantum machines improve in size and fidelity, the complexity of circuits executed on these machines will expand. The potential for mistakes and incorrect executions will increase, resulting in wastage of critical machine time and resources - we observe over 5\% wasted executions in our study. Thus, debugging and verification strategies are a must to maximize useful system utilization.}
    
    \circled{2}\ \emph{Compilation times are on the increase as we move to larger applications and machines - we observe more than a 1000x increase as we go from current day circuits to 1000q circuits. There is a need to build more scalable compilation strategies, identify compilation techniques which are optimally beneficial to the target circuit, as well as potentially overlap some compilation tasks with the already long queuing times.}

    \circled{3}\ \emph{Machine characteristics can vary widely across machines and time, and their impact on applications are often not  well-understood - we observe over a 3x fidelity variation across machines.  2-qubit gate based metrics are a reasonable indicator of an application's fidelity on a machine and can be evaluated at compile time. These metrics can then be used to aid in machine selection and users can be allowed to trade-off fidelity for low queuing time and vice-versa. }
    
    \circled{4}\ \emph{To maximize the overall system utilization / throughput and to improve application fidelity across users, opportunities for vendor-employed machine-aware system wide management of resources (with user-constraints) should be explored. But automated mechanisms will require more stability and homogeneity among available machines, a likely expectation for the future.}

    \circled{5}\ \emph{Lack of discipline in load distribution leads to very widely varying queuing times - we observe queuing times from under a minute to even over days. 
    Queuing time will become more challenging to predict as demand, supply and prioritization techniques continue to grow. Research on predicting queuing times are worth pursuing.}
    
    \circled{6}\ \emph{Long queuing times can not only reduce system throughput, but also reduce application fidelity - for example, by making device-aware compilations stale. 
    Dynamic circuit re-compilation based on machine monitoring is promising and would be particularly useful for the pulse-based compilation approach.}

    \circled{7}\ \emph{Execution times are considerably lower than queuing times (around 0.1x on average), even though there is variation across jobs and circuits.
    These variations are mostly influenced by machine characteristics rather than circuit characteristics.
    This is because the current complexity of NISQ-era quantum circuits are low enough that machine executions overheads are greater than the actual execution time of the circuit.}
    
    \circled{8}\ \emph{For the foreseeable future, execution times are likely to be highly predictable and mostly dependent on a few characteristics - we are able to predict executions times on most machines with over 95\% correlation. Predicting execution time accurately amplifies the possibility of efficiently implementing the recommendations related to scheduling, predicting queuing times, and leveraging queuing times in useful ways.}

%% file: 2_background.tex
\section{Background and Terminology}
\label{Background}
\subsection{Quantum Information}
\label{information-qc}

Quantum information science redefines the computational model through the use of quantum bits, or qubits, that have two basis states represented as  $\Ket{0} =  \begin{bmatrix}  1 & 0
\end{bmatrix}^\text{T}$ and $\Ket{1} =  \begin{bmatrix}  0 & 1
\end{bmatrix}^\text{T}$. Qubits, unlike classical bits that hold a static value of either 0 or 1, demonstrate states of superposition in the form of $\alpha\ket{0} + \beta\ket{1}$ with probability amplitudes $\alpha,\beta \in \mathbb{C}$ and hold values such that $|\alpha|^2+|\beta|^2=1$. Superposition enables $n$ qubits to represent up to $2^n$ states simultaneously, and this phenomenon, along with the ability for quantum states to interfere and become entangled, allow certain problems to be solved with significant reductions in complexity. Qubits hold large quantities of information for processing while in superposition, but upon measurement, quantum state collapses and only classical values of either 0 or 1 are observed.  

Some common single-qubit transformations include the $R_x(\pi)=X$, $R_z(\pi)=Z$, $R_y(\pi)=Y$ rotation operations that cause a bit flip, a phase flip, and a combination bit flip and phase flip, respectively. Additionally, the $H$ gate puts a qubit originally in a basis state into perfect superposition where it has equal likelihood of being measured as 0 or 1. Multi-qubit operation are critical for entanglement and examples include the logical $SWAP$ operation that causes the exchange of quantum state along with controlled gates, such as $CX$ or $CZ$  that execute an operation on a target qubit depending on the state of one (or more) control  qubit(s).

\label{NISQ-info}

Current quantum devices are extremely fragile, and as a result, some of the biggest challenges that limit scalability include limited coherence times, gate errors, readout errors and crosstalk.
In addition, most NISQ suffer from limited connectivity as configuration only permits nearest neighbor two-qubit execution.
%
The coherence times for superconducting quantum computers have improved from 1 nanosecond to 100 microseconds in the last decade and have recently targeted 1000 microseconds~\cite{Tannu:2019a}. 
Performing gate operations on qubits can also affect their state incorrectly due to errors. 
From public IBM information, single-qubit instruction error-rates are of the order of $10^{-3}$ , whereas for two-qubit instructions, such as CNOT, it is $10^{-2}$ (in terms of probability of occurrence)~\cite{Tannu:2019a}. 
Also, crosstalk arises from unwanted interactions between the qubits and from leakage of the control signals onto qubits which are not part of the intended gate operation~\cite{murali2020software}.

%
%
In this paper, we focus on IBM's superconducting circuits based quantum devices.
%
Our analysis has been performed on quantum jobs designed and submitted to the IBM quantum machines via IBM Qiskit~\cite{Qiskit}.

\subsection{Key Terminologies}
Below we provide terminology definitions, some of which are based specifically on the IBM Quantum Cloud but are generally applicable even otherwise.

    \circled{1}\ \textbf{\emph{Algorithm:}} 
    Describes quantum computation at the highest abstraction level.
    
    \circled{2}\  \textbf{\emph{Circuit:}} A single quantum circuit with a list of instructions bound to some registers. It has a number of gates and is spread out over a number of qubits. Gates are often 1-qubit and 2-qubit gates if the circuit is specified in low-level assembly code. 
    The width of the circuit is the number of qubits it requires and the depth of the circuit is the critical path, often counted as the number of 2-qubit gates in the critical path.
    
    \circled{3}\  \textbf{\emph{Compilation:}} Involves a sequence of steps to enable the quantum circuit to be executed on a specified quantum machine in a valid and efficient manner.
    
    \circled{4}\  \textbf{\emph{Job:}} Encapsulates a single circuit or a batch of circuits that execute on an quantum machine or simulator. The circuits within a batched job are treated as a single task such that all quantum circuits are executed successively. Further, each circuit in the job will be rapidly re-executed for a specified number of shots (eg. most IBM Q machines generally allow a batch size of 900 circuits with each circuit allowing up to 8192 shots). 
    
    \circled{5}\  \textbf{\emph{Queue:}} When a job is submitted to a quantum machine on the cloud, it enters a queue (for that particular machine) with jobs from other users before eventual execution. The order which these jobs are executed is, by default, determined by some fair sharing based queuing algorithm. 
    In practice, this means that jobs from various providers are inter-weaved in a non-trivial manner, and the order in which jobs complete is not necessarily the order in which they were submitted~\cite{IBMQE}. 
    
    \circled{6}\  \textbf{\emph{Results:}} Once a job is complete, the classical measurement results  from the job are returned to the client (as a count of bit-strings). These results are unique to each circuit in the job. For each circuit, the number of classical bit-strings returned equals the number of shots the circuit was executed for, and the width of the bit-string equals the number of data qubits in the circuit.

%% file: 3_obsA.tex

\section{Overall System Trends}
\label{OST}
In this section, we discuss overall system trends in terms of machine usage, time spent on compilation, queuing in the cloud and actual machine execution.

\subsection{Machine Executions}

\begin{figure}[t]
     \centering
     \begin{subfigure}[b]{0.67\columnwidth}
         \centering
         \includegraphics[width=\textwidth]{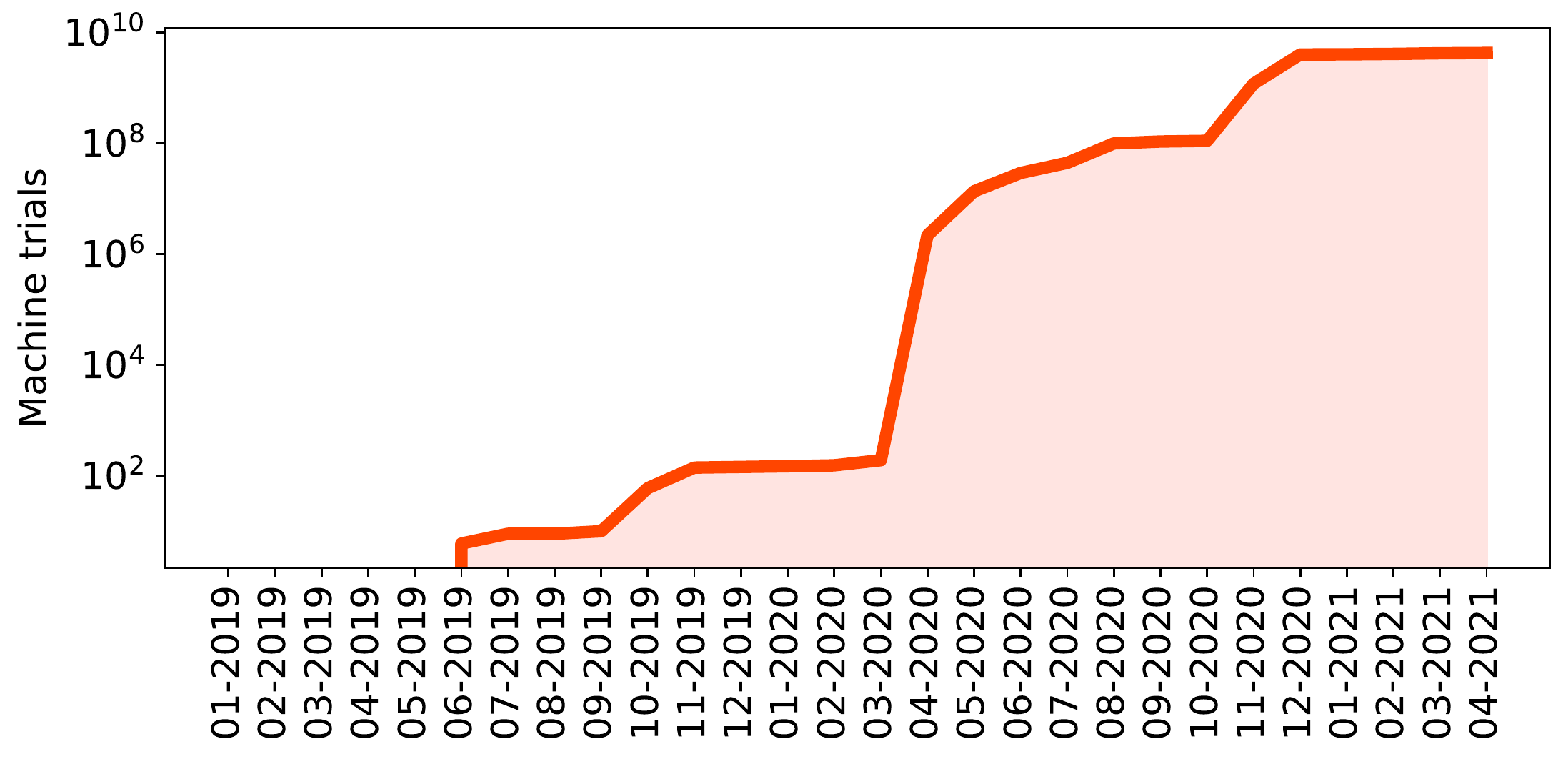}
         \caption{Machine Trials over 2 years}
         \label{Fig:SC21_Trials_Time}
     \end{subfigure}
     \begin{subfigure}[b]{0.31\columnwidth}
         \centering
         \includegraphics[width=\textwidth]{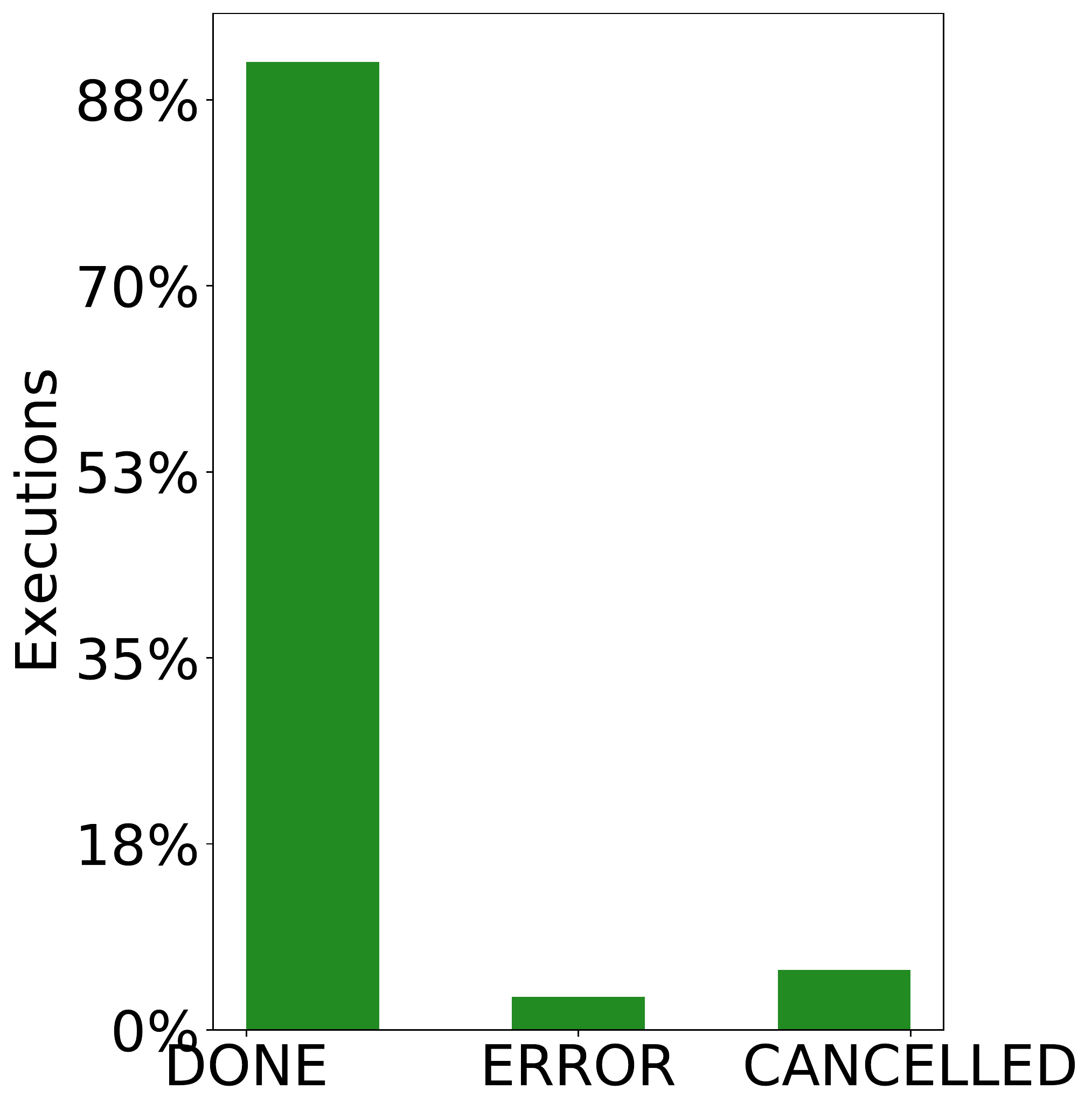}
         \caption{Status}
         \label{Fig:SC21_Correct}
     \end{subfigure}
        \caption{Cumulative quantum executions in our study and their validity. Executions have grown considerably over the past 12 months. Further, over 10\%  did not execute cleanly.}
\end{figure}

Fig.\ref{Fig:SC21_Trials_Time} shows a cumulative count of almost 10 billion executions / trials run on the quantum machines over a two year study period.
Shown on a log scale, it is evident that the number of jobs being run on the machines are consistently increasing at growing pace as the popularity of quantum computing continues to grow.
This is coupled with an increase in the number of quantum machines supporting a larger number of qubits and with better fidelity, meaning that the scope of the experiments run / problems targeted on the quantum cloud continue to steadily grow.
We also note that the growth in quantum usage is considerably greater than classical super-computing usage.
Recent work analyzing supercomputing usage trends~\cite{Patel:2020} has shown that the number of jobs halved over a 10 year period, while the size of jobs grew by 7x. 

This effective 3.5x growth in supercomputer usage over a decade is expected to be small in comparison to the envisioned growth in quantum computing usage over the coming decade - this is also indicated by trends seen in our sample study space.

This is an expected trend with new technologies - there will be an exponential growth in resource requirements in the coming decade.

In Fig.3.b, we show a breakdown of the status of execution of the quantum jobs on the machines.
While around 95\% of the jobs were successfully executed, around 5\% errored out or were cancelled.
Note that successful execution here means that the jobs were executed to completion on the machine - it is not indicative of the quality of the result of the quantum execution.
Considering that the cost of executing quantum jobs (both in terms of time and money) is expensive in the NISQ-era, it is important to maximize job success rates.
This will be more challenging as quantum algorithms steadily grow in complexity.

\subsection{Queuing Time}

\begin{figure}[t]
\includegraphics[width=\columnwidth,trim={0cm 0cm 0cm 0cm},clip]{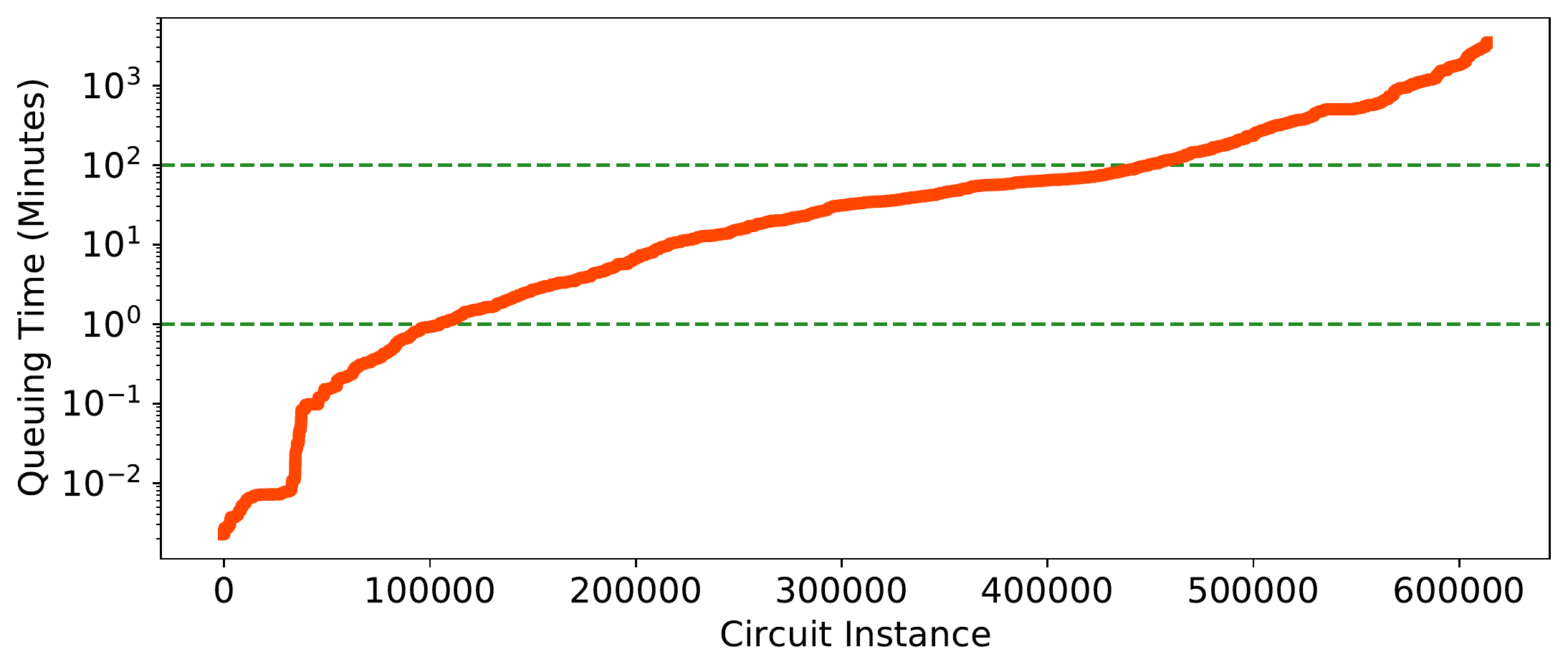}
\centering
\caption{Queuing time experienced by circuits run on the IBM Quantum machines (sorted)  over two years. Green lines correspond to times of 1 minute and 2 hours respectively. Jobs shown are a mix of public and privileged jobs. A considerable fraction of jobs experience large queuing times.} 
\label{Fig:SC21_QT}
\end{figure}

The fast paced growth in quantum computing requirements have resulted in the contention trends for these already scarce resources to continually worsen.
And a first-order impact of quantum machine scarcity are the long queuing times experienced while accessing cloud machines. 

Fig.\ref{Fig:SC21_QT} plots the cloud queuing time experienced by the executed circuits in our study, in a sorted order.
Note that these executed circuits are through a mix of public and privileged (i.e. paid) access to these quantum machines.
Also, fair-share queuing executes jobs on a quantum system in a dynamic order so that no user can monopolize the system~\cite{IBM-Fair} - more details can be found at the source.
Only around 20\% of the total circuits experience ideal queuing times of, say, less than a minute.
The median queuing time is around 60 minutes which is not insignificant.
Further, more than 30\% of the jobs experienced queuing times of greater than 2 hours, and around 10\% of the jobs were queued up for as long as a day or even longer!
The classical HPC systems analyzed in \cite{Patel:2020} estimated that the average queuing times on their supercomputers increase from 0.1 hours to 1.2 hours over a decade.
The current queuing times for quantum clouds, even at this stage of relative infancy, are already comparable to the higher side of the classical queuing times.
A similar 10x increase in quantum waiting times over the next decade would be detrimental to quantum research and development.
The higher queuing times are especially concerning, considering that the actual quantum execution runtime on the quantum machines is only in the order of seconds or minutes - this is discussed next.

\subsection{Execution vs Queuing Time}

In comparison to the earlier analysis of queuing time, the execution times are far lower, with over 99\% of the circuits executing in less than 0.1 minutes.
Note that these numbers are per-circuit. 
As described earlier, a quantum job can be made up of a batch of a number of circuits.
This means that the execution time for a job is usually at or under an hour, even with full batch utilization. 
In comparison, \cite{Patel:2020} shows that the median HPC runtime in 2018 was just over an hour, with a 60\% increase over a decade.
Thus, it is intuitive that less than 100\% batch utilization would make the relative waiting time much worse in the quantum scenario compared to the classical HPC setting. 
In the future, as application circuits become larger it is still unclear how the execution time will grow - in Section \ref{ExePred}, the relation between execution time and different job / machine characteristics are discussed.
However, it is very likely that queuing times grow faster than execution times.
Queuing to execution ratio is discussed next.


\begin{figure}[t]
\includegraphics[width=\columnwidth,trim={0cm 0cm 0cm 0cm},clip]{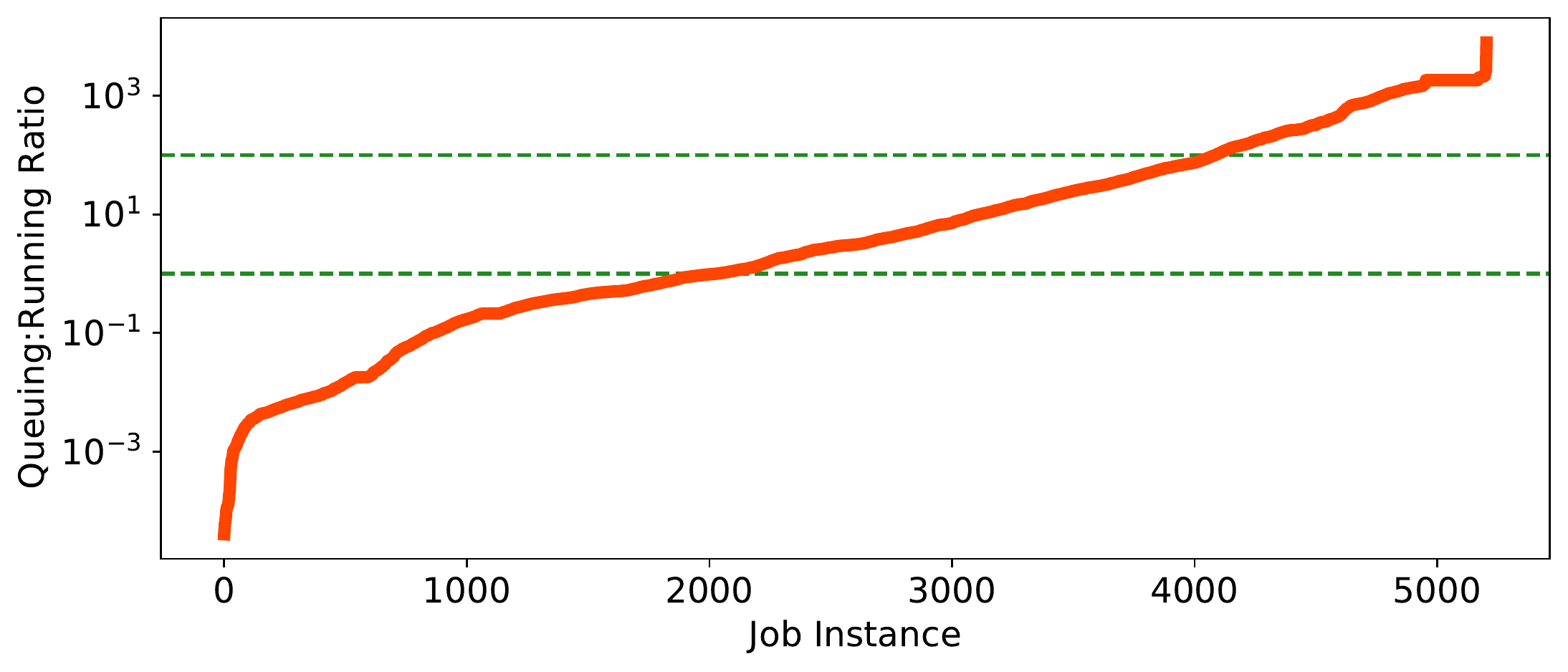}
\centering
\caption{Ratio of queuing times to execution times (sorted).Green lines correspond to ratios of 1x and 100x respectively. The median ratio is around 10x.} 
\label{Fig:SC21_QR}
\end{figure}

Fig.\ref{Fig:SC21_QR} shows per-job queuing to execution time ratios in a sorted order.
In around 30\% of the total quantum jobs, the experienced queuing time is at par or lower than the execution time, which is the ideal scenario.
These are usually jobs which are a) maximizing batch utilization by batching close to 100\% of the possible circuits into a single job, and b) running on machines with lower number of jobs lined up.
On the other hand, the median ratio is around 10x and around 25\% of the total jobs experience ratios which are 100x or more. 
Meticulous management of jobs and resources is required as contention continues to drastically increase (at least until the supply can sufficiently cater to the demand).

\subsection{Compile Time}

\begin{figure}[t]
\includegraphics[width=\columnwidth,trim={0cm 0cm 0cm 0cm},clip]{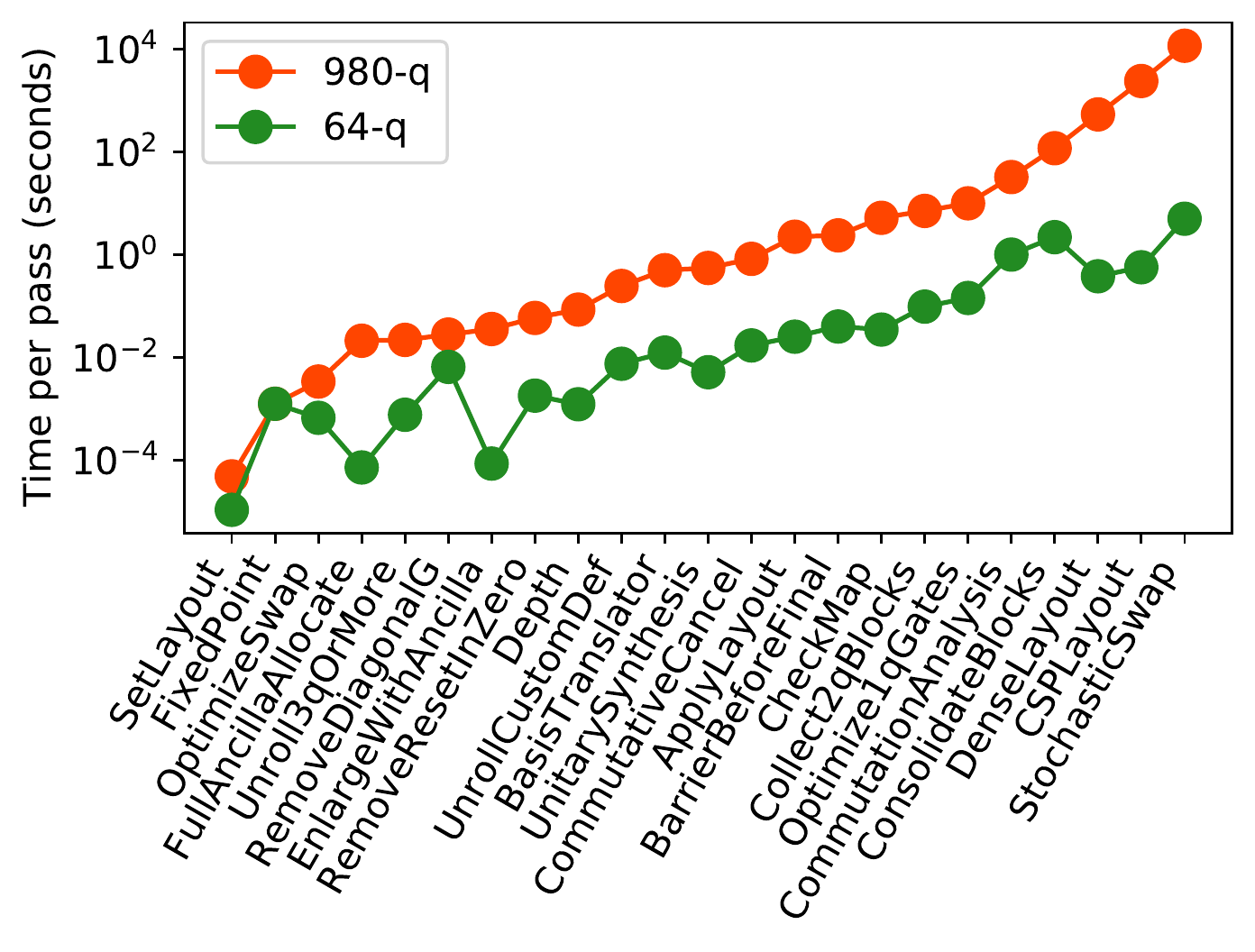}
\centering
\caption{Time per pass across different passes employed by Qiskit at its highest compiler optimization level. The layout and routing passes are especially expensive. 
} 
\label{Fig:PassTime-QFT}
\end{figure}

Next, we analyze compilation times for quantum circuits.
While compilation is traditionally independent of the cloud, we believe it is critical to consider all of these times/overheads in unison, so as to propose effective solutions for maximizing user application fidelity and system throughput.

Fig.\ref{Fig:PassTime-QFT} shows quantum compilation for a 64q Quantum Fourier Transform (QFT) circuit targeting a 65-qubit quantum machine (IBM Manhattan) and an illustrative 980q QFT compiled for a fake 1000-qubit quantum machine.
In the 64-qubit case, which is somewhat an upper limit for current quantum computers, compilation times are in the order of roughly one second or less for all compiler passes meaning that the entire compilation process can be completed in minutes.
On the other hand, compilation times drastically scale up by 100-1000x as we compile for 1000 qubits, a potential target in the near future.
The layout and routing passes are especially expensive, potentially consuming more than 10k seconds (~3 hours). 
These numbers will continue to grow drastically with the current state-of-the-art compilation techniques.

Further, unlike the classical world, optimal quantum compilations are dependent on dynamic (or at least, current) machine characteristics.
Thus, for compilations to be close to optimal, they should be performed on the most recent state of the machine prior to actual execution - this makes the impact of long compile times even worse.
Increasing compilation times can especially impact throughput at the user end in the context of iterative applications where a circuit can be built (and thus compiled) only after processing data/feedback received from prior job executions on the quantum machine.

Orthogonal to the results presented above, prior work~\cite{Shi_2019} discusses that compilation to the pulse level (a critical direction for efficient quantum computing) can also be very cumbersome and can consume several hours of compilation time.
Similarly, search algorithms for approximate quantum circuit synthesis can take minutes to hours - prior work~\cite{younis2020qfast} has shown that 6-qubit circuits can take as much as 3-4 hours for synthesis and these numbers would scale up with larger circuits.


\subsection{Summary and Recommendations}

    \circled{1}\  \emph{The complexity of circuits executed on quantum machines will increase as machines get larger and error rates improve. 
    This means that the potential for mistakes and incorrect executions will increase, resulting in wastage of critical machine time and resources.
    Thus debugging and verification strategies~\cite{Huang:2019,Rand:2019} in quantum computing are a must for maximization of useful system utilization.
    Checks can be employed at both the user-end as well as by the vendor.}
    
    \circled{2}\  \emph{Compilation times are also on the increase. As we target circuits with 1000 qubits or more, compilation times can run into many hours - becoming unsuitable for dynamic machine compilation. There is a need to build more scalable compilation strategies, identify compilation techniques and appropriate optimization thresholds which are optimally beneficial to the circuit at hand, as well as potentially overlap some of the time consuming compilation tasks with the already long queuing times.
    Compilation passes can be separated into those that are minimal requirements and those which are nice-to-have optimizations. 
    Passes can be implemented, if possible, as progressive optimization algorithms.}
    
    \circled{3}\  \emph{ With the increasing popularity of quantum computing, the number of jobs executed on quantum machines will continue to grow drastically and thus optimizing the available resources at both the vendor-end and the client-end is critical for maximizing system effectiveness and throughput.}
    
    \circled{4}\  \emph{Increase contention for cloud resources means increase waiting time for their access. From observing HPC analyses, it is plausible that quantum queuing times increase by 10x or more if current trends are maintained. This could mean average waiting times of half a day or more. Thus, it is critical to a) reduce wait times by better management of resources and b) potentially recycle waiting time by performing suitable tasks while the jobs are queued, that can improve execution efficient for the user as well as for the system as a whole. }
    
    \circled{5}\  \emph{ While execution times might also be expected to increase, the queuing to execution ratios are drastically high and are likely to keep increasing in the quantum world.
    User-employed efficient batching of circuits into jobs via better knowledge / understanding of the applications at hand, can alleviate this to some extent. }

%% file: 4_obsB.tex

\section{Trends in Quantum Machines}
\label{machines}
In this section, we analyze some variations across the different quantum machines accessed on the cloud and their potential impacts.
Our study encompasses 25 different quantum machines with qubits ranging from 1 to 65. 

\subsection{Variable Machine Characteristics}

\begin{figure}[t]
\includegraphics[width=\columnwidth,trim={0cm 0cm 0cm 0cm},clip]{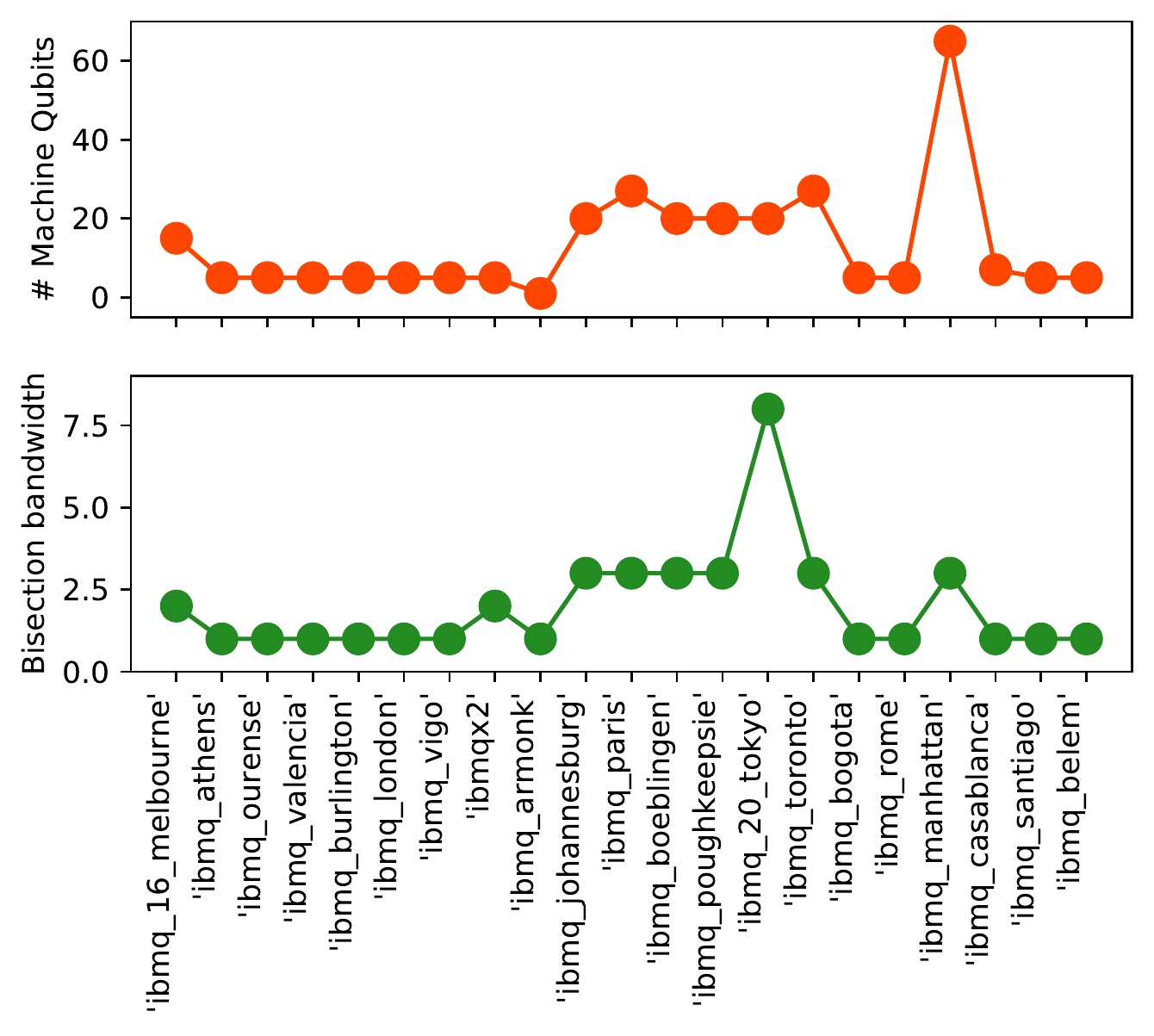}
\centering
\caption{Qubits vs Bisection Bandwidth. In contrast to common classical computing systems, the bisection bandwidth is very low across these quantum machines.} 
\label{Fig:SC21_Machine}
\end{figure}

Different machines experience different characteristics stemming from their topology, qubit strength, quality of calibration and drift. 
Fig.\ref{Fig:SC21_Machine} shows the qubits and bisection bandwidth across the different quantum machines.
Bisection bandwidth is a common metric in classical computer networking - if the network is bisected into two partitions, the bisection bandwidth of a network topology is the bandwidth available between the two partitions.
Bisection bandwidth is an accurate metric to measure the connectivity of a topology.

In contrast to common classical computing systems, it is evident from the figure that the bisection bandwidth is very low across these quantum machines, especially so in the larger machines.
For example, the 65 qubit IBM Manhattan has a bisection bandwidth of 3. 
In comparison, a 64-node classical system employing a standard mesh topology would have a bisection bandwidth of 8.
The low connectivity in these machines can be attributed to the high noise levels in these machines especially when there is increased proximity/coupling among gates/qubits (eg. crosstalk effects~\cite{Murali_2020}).
This restricts the ability to efficiently run larger applications even on larger quantum machines, apart from other dynamically varying noise characteristics and so forth.
This also makes user choices harder, to decide which machine is most suited to run their target application.
It will be interesting to see how these trends shape up in the near future: on the one hand, qubits are becoming more robust, but on the other hand, as machines with more qubits are built, any resultant increasing qubit / gate density can worsen some specific noise characteristics. 

\subsection{Varying Application Fidelity}
\label{VAF}

Characteristics like the above result in potentially different circuit fidelity (or "Probability of Success") across different machines and these characteristics can also change over time.
NISQ-era quantum machines are affected by non-deterministic spatial and temporal variations in their characteristics, for instance, in terms of their one- and two-qubit error rates.
Prior work~\cite{Tannu:2019a} studied a 20-qubit IBM machine for 52 days and observed the prevalence of a wide distribution of machine characteristics with considerable spatial and temporal variation.
From the spatial perspective, they observe the coefficient of variation (i.e. ratio of the standard deviation to the mean) to be in the range of 30-40\% for $T_1$/$T_2$ coherence times, as well as nearly 75\% for 2-qubit error rates - clearly indicative of wide variation across the machine.
From a temporal perspective, they observe more than 2x variation in error rates in terms of day-to-day averages - these variations are impacted by both day-to-day calibration of these machines, as well as drift between calibrations.

\begin{figure}[t]
\includegraphics[width=\columnwidth,trim={0cm 0cm 0cm 0cm},clip]{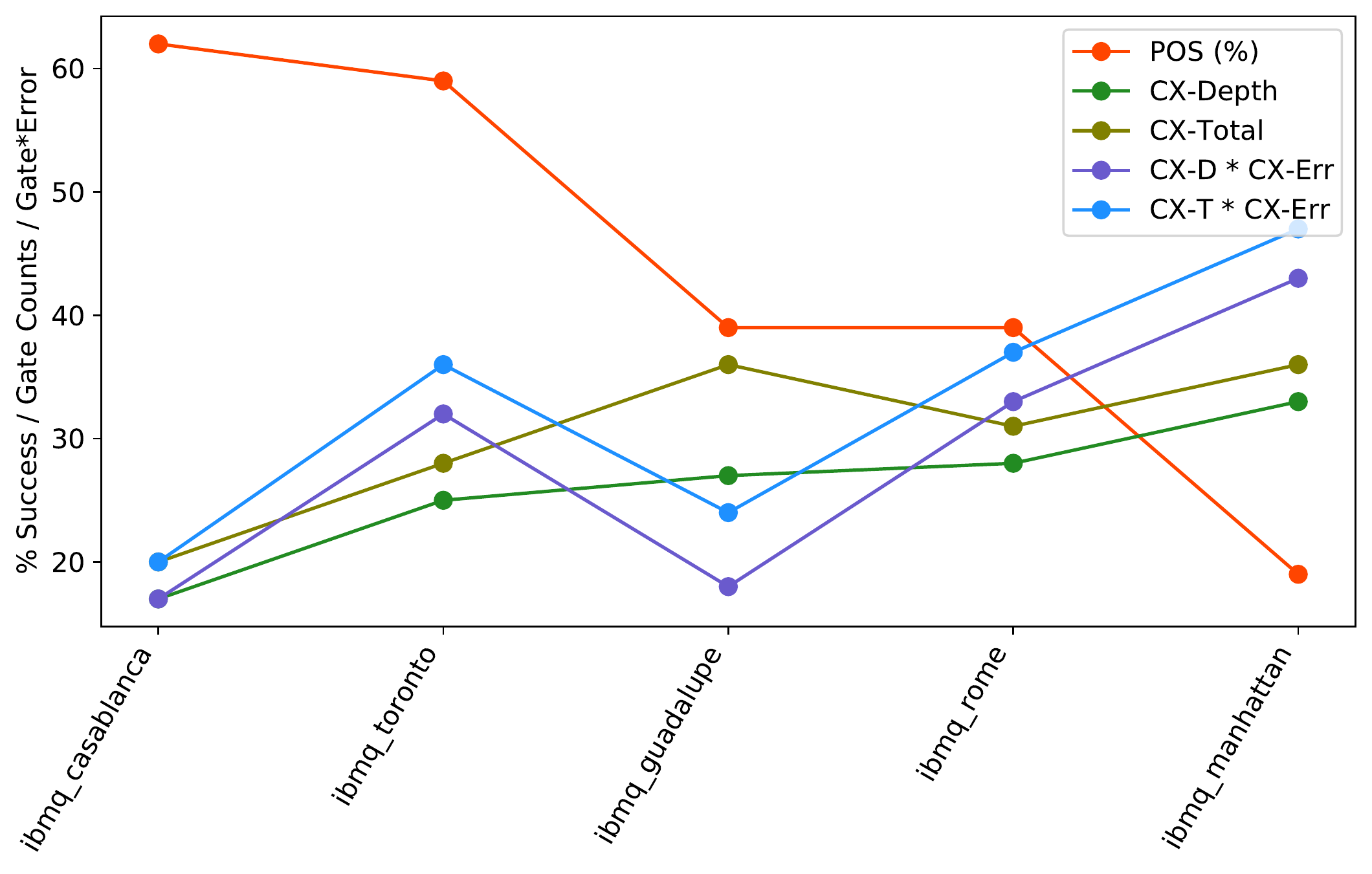}
\centering
\caption{Fidelity of 4q QFT vs CX metrics (CX-Depth, CX-Total, CX-D * CX-Err and CX-T * CX-Err) over multiple machines. Fidelity correlates with CX characteristics.} 
\label{Fig:SC21_Spatial}
\end{figure}

Users often select an optimal machine to run on based on its known characteristics and suitability to a particular application at hand - but this task can be complicated.
This is shown through Fig.\ref{Fig:SC21_Spatial}, which presents results from evaluating a benchmark application (4-qubit Quantum Fourier Transform) across a set of IBM Quantum machines - Casablanca (7q), Toronto (27q), Guadalupe (16q), Rome (5q) and Manhattan (65q).
Observe that the Probability of Success (POS), in orange, can vary widely across these machines - varying from 62\% success to 19\%.

First, note that the POS is not directly correlated with the size of the machine.
The  highest POS is observed on a 7q machine while the lowest is on the 65q machine.
Naively, it would be expected that any application would likely be more suited to larger quantum machines because a larger machines has a greater number of qubits to choose from (i.e. to map the application to the best possible qubits).
This highlights that a better unified understanding of machine/technology generation, topological constraints as well as dynamically variable noise characteristics is required.

Second, the figure shows that the POS is well correlated with two-qubit gate (CX) characteristics.
In figure, we show 4 metrics related to the CX-gate on the target machine - a) CX-Depth: the depth of the circuit's critical path in terms of CX gates, b) CX-Total: the total number of CX gates in the circuit, c) CX-D * CX-Err: the depth of the circuit multiplied by average error of the circuit's CX gates and d) CX-T * CX-Err: same as (c) but with total CX gates.
It is evident that the POS of success decreases as the CX-metrics tend to increase.
The CX gate is critical to application fidelity because it surpasses 1-qubit operations in terms of both gate error and execution time on superconducting hardware~\cite{jurcevic2021demonstration}.
Greater the number of influential CX gates potentially implies lower circuit fidelity.
Thus, analyzing the CX-metrics from an application after its compilation for a machine is a potential indicator of the fidelity on that machine.



Fidelity is also a concern in the classical computing / HPC cloud for specific classes of applications like machine learning.
Fidelity is dependent on machine related characteristics - which ML models are employed, what bit precision is used, how large is the data set, how long is the training pursued and so on.
But these dependencies are deterministic in comparison to those in the quantum computing world wherein characteristics change more rapidly/significantly and in ways that are yet to be well understood.

\subsection{Machine Utilization Distribution}

\begin{figure}[t]
\includegraphics[width=\columnwidth,trim={0cm 0cm 0cm 0cm},clip]{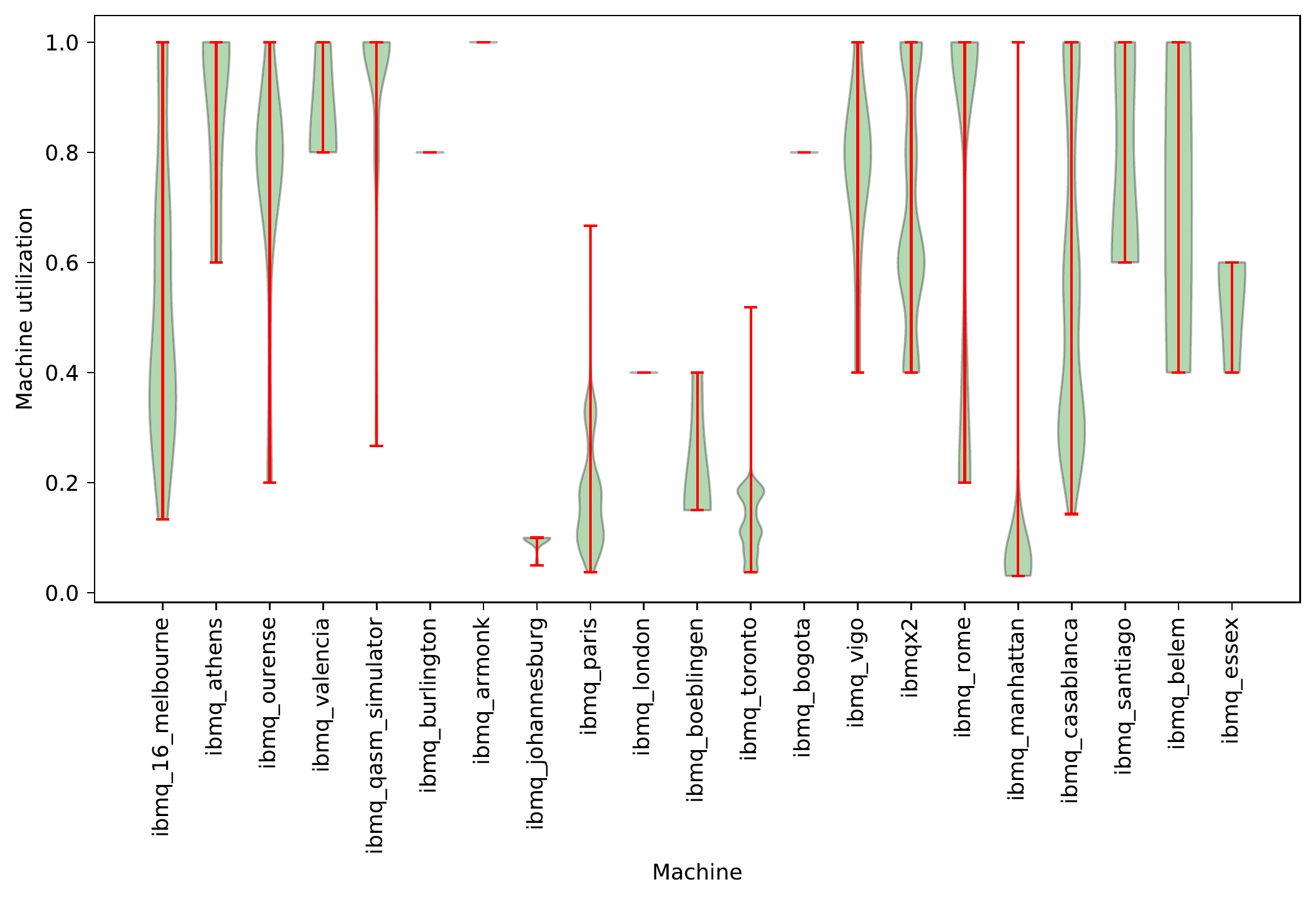}
\centering
\caption{Machine Utilization by circuits vs Machine, where utilization is defined as the fraction of machine qubits that were used by the circuit.} 
\label{Fig:SC21_Machine_Util}
\end{figure}

Fig.\ref{Fig:SC21_Machine_Util} shows a violin plot of the machine utilization of the circuits run on each quantum machine, where utilization is defined as the fraction of machine qubits that were used by the circuit. 

First, it is evident that while the utilization is higher on the smaller machines, it is lower on the larger ones.
It is intuitive that utilization is lower on the larger machines for reasons such as connectivity and bisection bandwidth discussed prior. 
Larger applications (i.e. requiring more qubits) are challenging to run due to connectivity constraints. 
These constraints increase the circuit depth (by inserting SWAP operations) which can then lower fidelity due to limited coherence times. 

Second, utilization is not uniform even among machines of same size.
It is not uncommon that some of these machine characteristics are unknown or minimally understood by the end user, and choices are often made after experimental evaluations which could waste both resources (machine usage) and time (compilation + queuing + execution).
Choice of machine is often based on characteristics of the machine at the point of use, such as gate error rates, coherence times etc, which can vary across calibration (i.e. when machine is retuned) cycles - note that it is often the case that such user decisions are driven by heuristics and not thorough hypotheses.

Third, choice of machine is also critically based on how busy particular machines are at the point of use - some machines are significantly more queued up than others (more in Section \ref{Queuing}), meaning that a sub-optimal machine can be chosen for quicker turnaround.

Improving utilization is actively explored in HPC.
Methods have been employed in parallel supercomputers when the job is moldable and resources are allocated based on system optimality~\cite{1178881}.

\subsection{Summary and Recommendations}

    \circled{1}\  \emph{Machine characteristics can vary widely across machines and over time on each machine. Moreover, many characteristics such as those related to topology, effects of noise, gate interactions etc. and their impact on a particular application might not be known to or likely not be understood well by users.  To alleviate this to some degree, our analysis shows that CX-gate based metrics are a reasonable indicator of an application's fidelity on a machine and can be evaluated at compile time. These metrics are influenced by both topology and noise characteristics and can thus be used to aid in machine selection.}
    
    \circled{2}\  \emph{In order to maximize the overall system utilization and throughput and to improve application fidelity across users, opportunities for vendor-employed machine-aware system wide management of resources should be explored, inspired by HPC. Vendor-managed allocation could be possible from within a set of machines chosen by a user (or all machines). 
    But note that automated mechanisms will require more stability and homogeneity among available machines, a likely expectation for the future.}
    
    \circled{3}\  \emph{While running larger applications might be restricted on even larger machines due to topology constrains, there is opportunity to improve machine utilization by multi-programming on the quantum machines~\cite{das:2019} i.e. running  multiple applications in conjunction on the machines. Choice of applications to run in conjunction could be influenced by system load, application fidelity requirements, dynamic machine characteristics etc.}

%% file: 5_obsC.tex

\section{Trends in Queuing}
\label{Queuing}

In this section, we analyze the queuing trends across different machines, batch sizes and also examine quantum specific detrimental effects of long queuing times.

\subsection{Number of Queued Jobs vs Machine}
\begin{figure}[t]
\includegraphics[width=0.95\columnwidth,trim={0cm 0cm 0cm 0cm},clip]{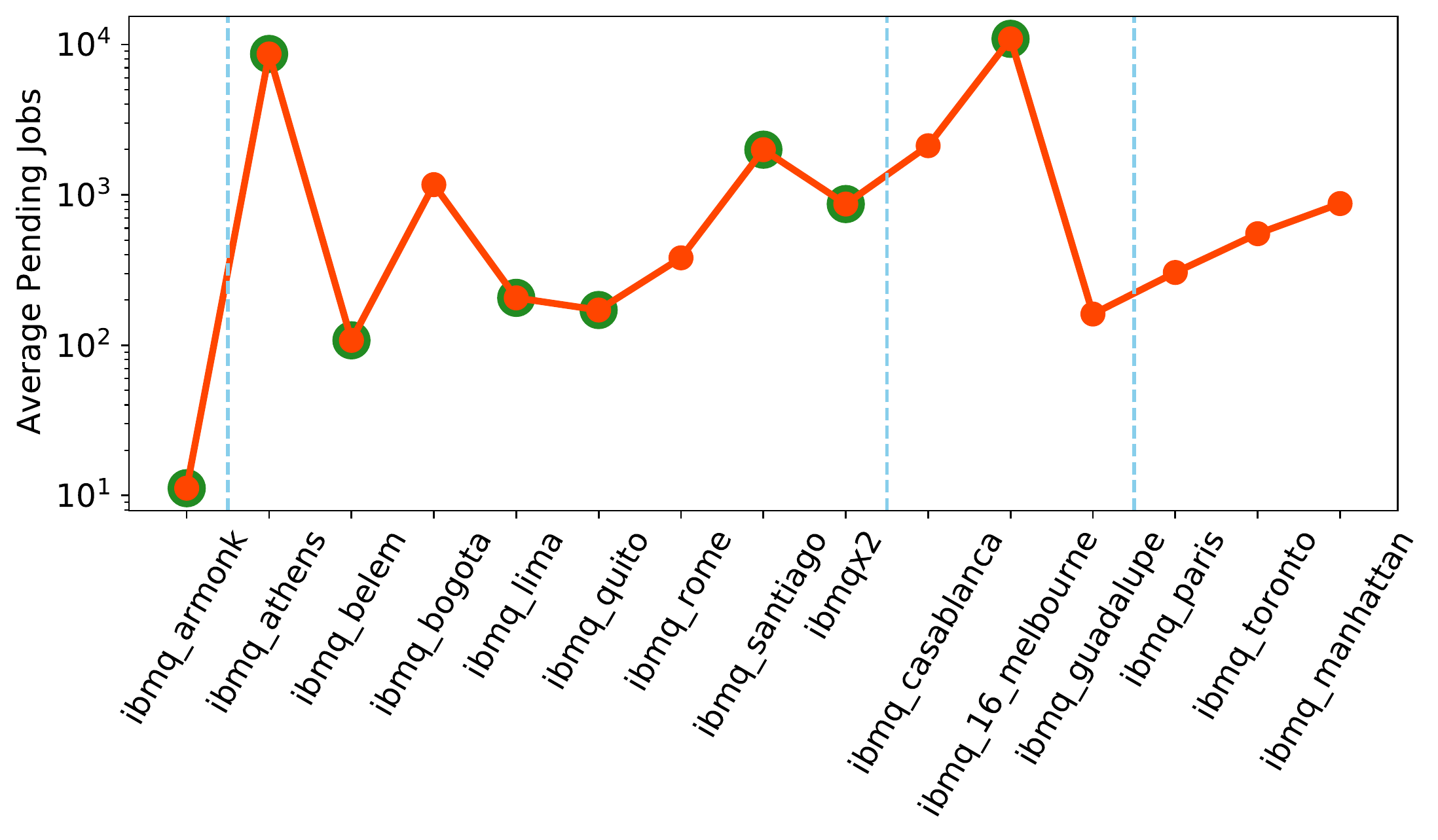}
\centering
\caption{Average pending jobs across different quantum machines, averaged over a week's period in March 2021. Jobs are unequally distributed across similar machines.} 
\label{Fig:SC21_MachineQJ}
\end{figure}

Fig.\ref{Fig:SC21_MachineQJ} shows the number of pending jobs across different quantum machines, averaged over a week's period in March 2021.
The machines are broken down into blocks (blue dashed lines) based on the number of qubits in the machine.
The first block is a 1-qubit machine, the next block is 5-qubit machines, the next is 7-16 qubits and the final is 27-65 qubits.
Further, publicly accessible machines are highlighted in green.
In each block, it is observed that the average pending jobs are highest on a public machine - this is expected since public machines have considerably more demand.
For instance, IBMQ Athens is 10-100x more in demand than other 5-qubit machines.
It is also observable that jobs are not distributed equally across machines (public or otherwise).
This can be associated with the characteristics of the machine and user-defined heuristics based on these characteristics (as discussed in Section \ref{machines}).
Note that while specifics of machine usage and machine popularity might change over time, the trends that jobs are unequally distributed across machines and that public machines are considerably in higher demand are expected to be consistent. Also note that all other data presented in this analysis are all collected across a two year period.

\subsection{Queuing Time Distribution vs. Machine}

\begin{figure}[t]
\includegraphics[width=\columnwidth,trim={0cm 0cm 0cm 0cm},clip]{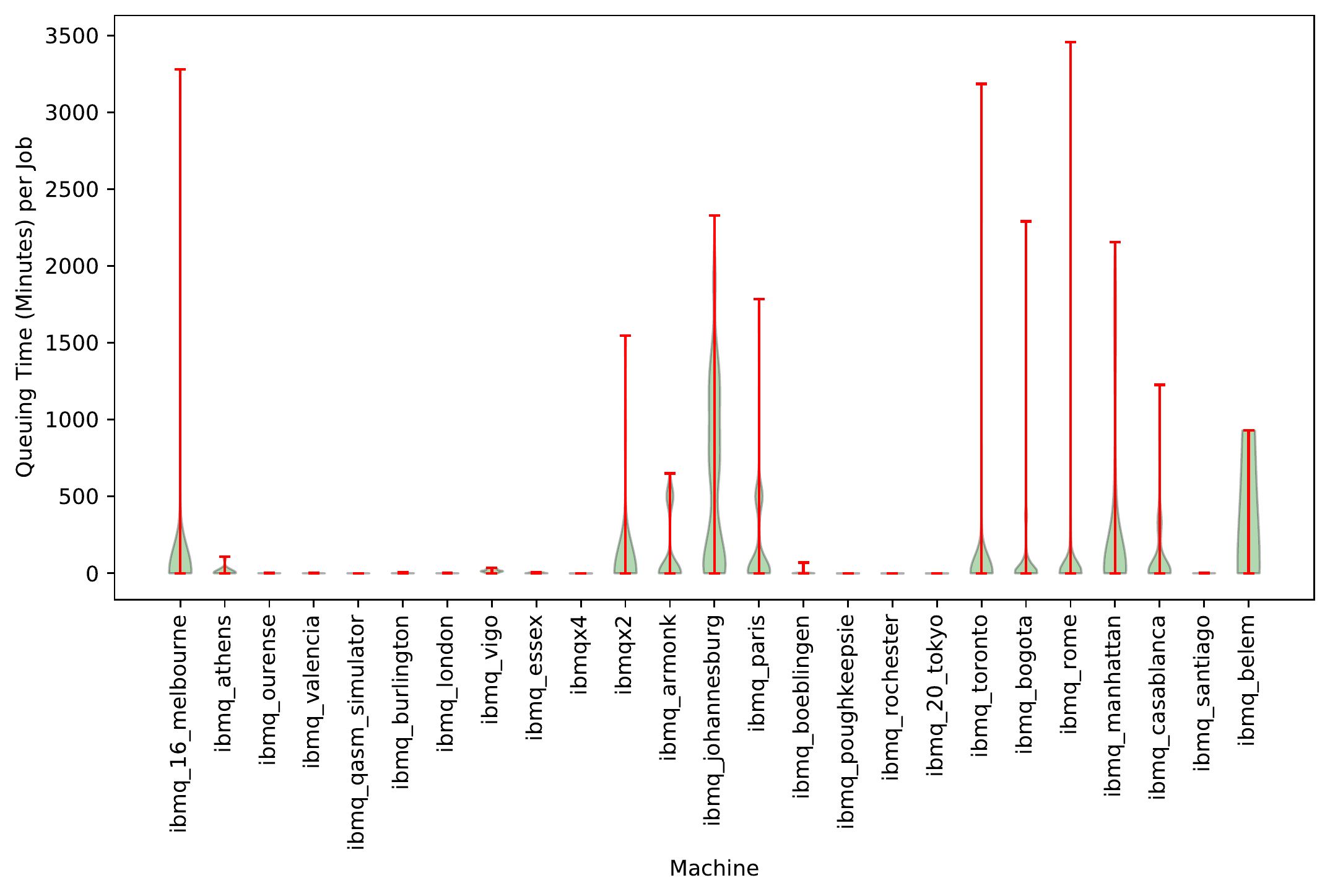}
\centering
\caption{Queuing time distribution of jobs vs Machine, over the two year period. On public access machines, the mean queuing times are of the order of multiple hours.} 
\label{Fig:SC21_QT_Machine}
\end{figure}

In Fig.\ref{Fig:SC21_QT_Machine} we show a violin plot distribution of queuing times per job across these quantum machines from over the two year period.
Queuing times can vary widely across machines - more than half the machines have queuing times varying from a few minutes to longer than a day.
On public access machines, the mean queuing times are of the order of multiple hours.
Privileged access machines, especially those with more qubits, can also often have high demand resulting in queuing times averages around a couple of hours, while they are usually one hour or less on other machines.

Queuing times also vary drastically in the classical compute space.
In fact, predicting queuing times is a hard to solve problem and has been the focus of extensive research~\cite{brevik-2006,hariharan2020endtoend}.
The interaction between workload and queuing discipline makes the amount of time a given job will wait highly variable and difficult to predict~\cite{brevik-2006}.
The quantum cloud is still at the nascent stage and strategies for queuing and job scheduling are simplistic at the present. 
Moreover, execution times of quantum circuits are relatively homogeneous and easier to predict - discussed further in Section \ref{ExePred}.
This makes queuing time prediction in quantum clouds less challenging for the time being.
But as the quantum supply and demand grow, it is fair to expect that queuing time prediction challenges from the classical world will be felt in the quantum cloud.

\subsection{Queuing Time Distribution vs Batch Size}
\label{QTBatch}
\begin{figure}[t]
\includegraphics[width=\columnwidth,trim={0cm 0cm 0cm 0cm},clip]{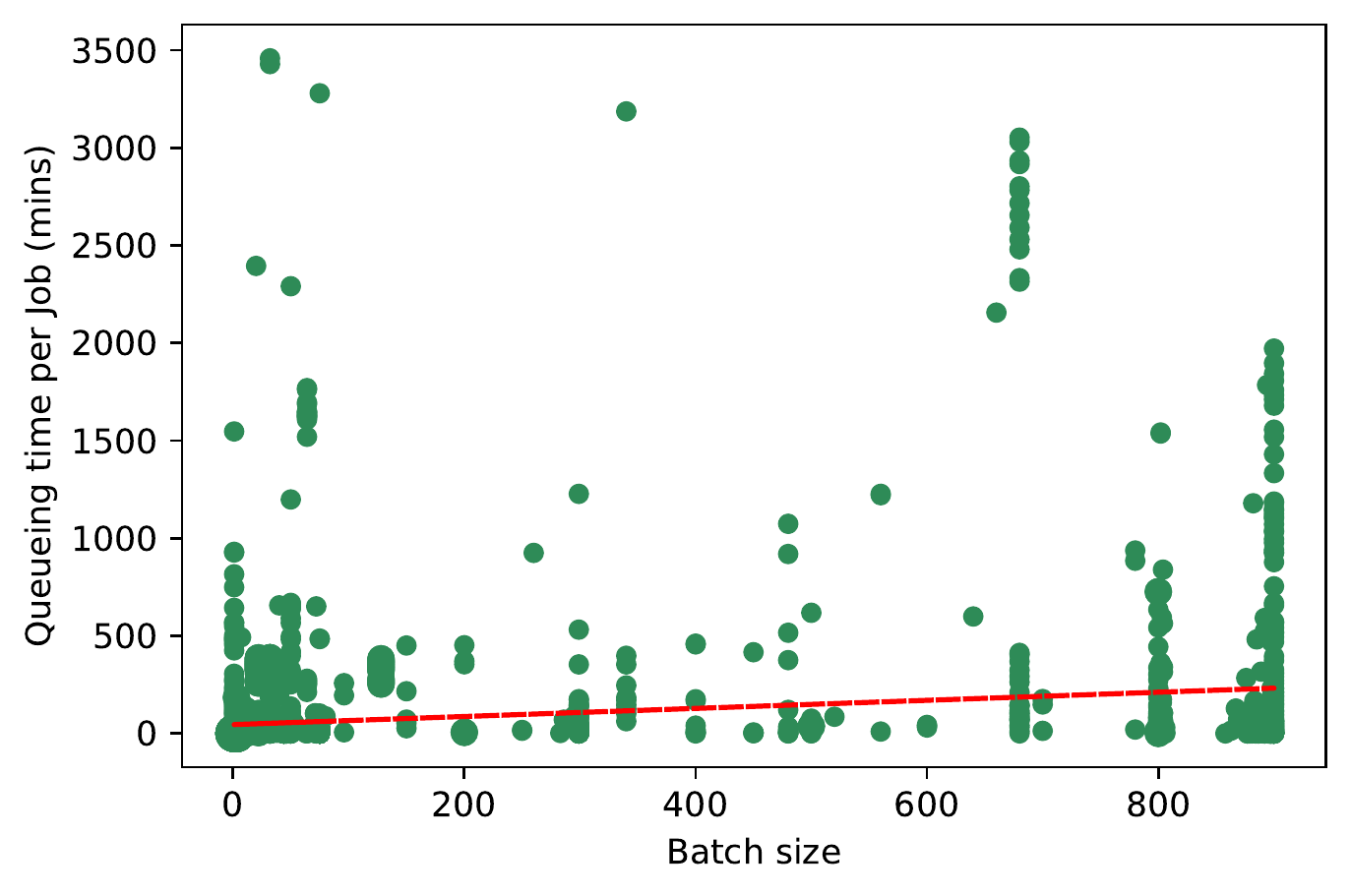}
\centering
\caption{Queuing time distribution of jobs vs batch size. As batch sizes increase, the effective queuing time \emph{per circuit} almost always decreases.} 
\label{Fig:SC21_QT_Batch}
\end{figure}

In Fig.\ref{Fig:SC21_QT_Batch} we show a comparison of queuing times and batch size of the job submitted.
Note that a job is made up of a batch of circuits - the limiting size in our experience is 900 circuits per job.

First, the figure shows that there is a wide distribution of batch sizes varying from 1-900.
It can be inferred from the above that it is often the case that a user's target application does not require the entire batch size and/or that users are not always adept at combining their executed circuits into a highly batched job. 

Second, as batch sizes increase, the queuing time \emph{per job} increases.
Some increase is expected because this particular job would take a longer time to execute, and thus increase the waiting time of future jobs.
This would especially be reflected if there are fewer jobs on a machine i.e. each jobs execution time has a larger influence on the queuing time on the machine.

Third, as batch sizes increase, the effective queuing time \emph{per circuit} almost always decreases.
This is because the entire batch of circuits is executed back to back in sequence and only suffer the queuing time once, as a whole.

Batching is an important area of research in Machine Learning and Distributed Computing.
Larger batches increase response time, but make better use of available network / memory bandwidth.
Similar trade-offs will exist in the quantum space, especially in the context of iterative quantum applications.

\subsection{Calibration Crossovers}
\begin{figure}[t]
     \centering
     \begin{subfigure}[b]{0.48\columnwidth}
         \centering
         \includegraphics[width=\textwidth]{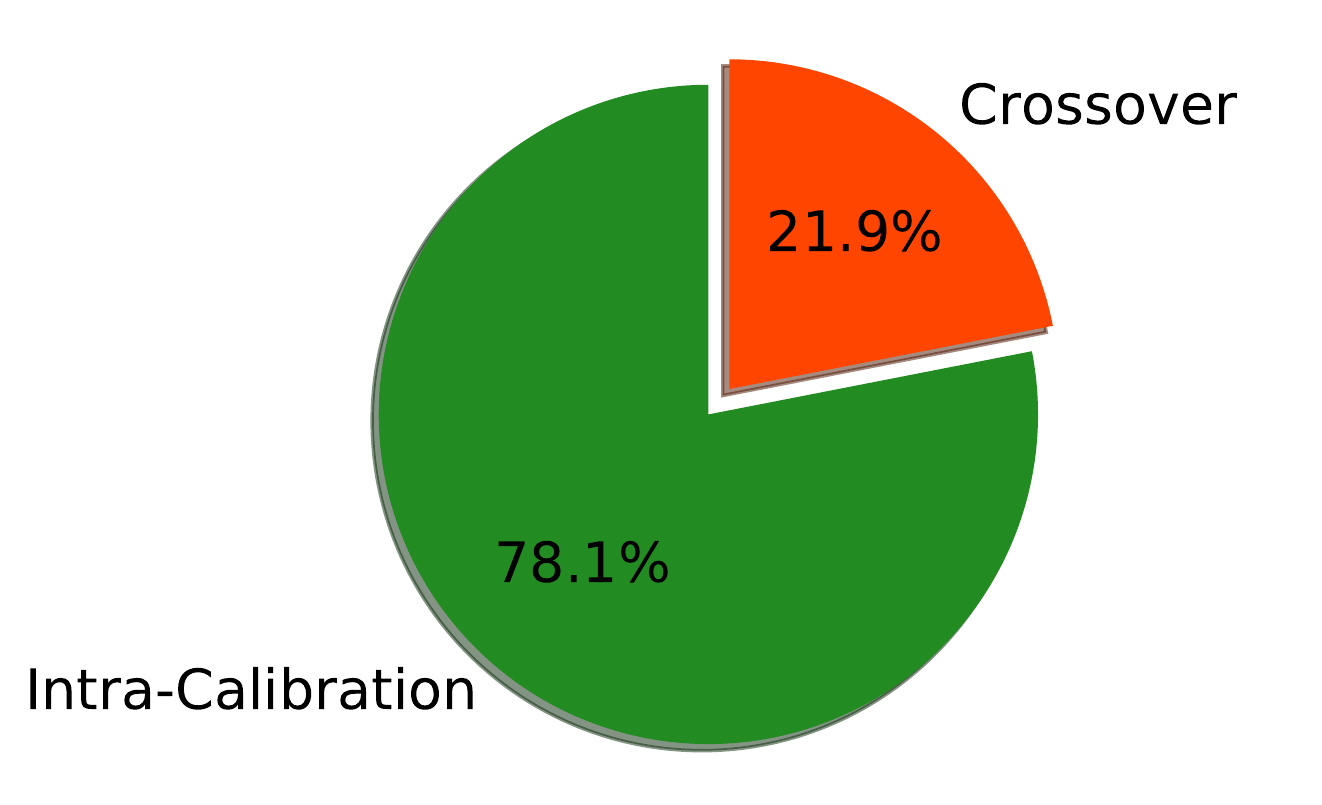}
         \caption{Jobs crossing calibrations}
         \label{Fig:SC21_Cross1}
     \end{subfigure}
     \begin{subfigure}[b]{0.5\columnwidth}
         \centering
         \includegraphics[width=\textwidth]{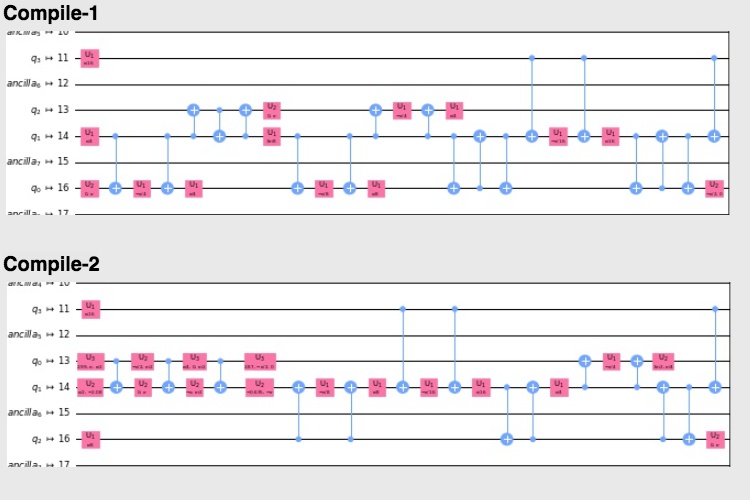}
         \caption{Varying Compiled Circuits}
         \label{Fig:SC21_Cross2}
     \end{subfigure}
        \caption{Effects of calibration crossovers. QCs can become sub-optimal over time.}
\end{figure}

When quantum circuits are compiled, they are done so in a device aware manner. 
While this involves static characteristics such as device topology and device basis gates, it also involves incorporating dynamic characteristics such as gate / qubit fidelity. 
As discussed earlier, the latter are dynamic because they evolve over time - these characteristics of qubits and gates are re-calibrated at some coarser granularity (say once a day) and these calibrations are non-uniform i.e. one day's qubit fidelity can be very different from the next day's qubit fidelity. 
Further, these characteristics also drift over time - meaning that they can differ even within a single calibrated epoch. 

Thus it is often the case that in scenarios of long queuing times, the dynamic characteristics which are accounted for at the  time of compilation are very different from the dynamic characteristics of the quantum machines at the time when the quantum circuit is actually executed on the machine. 
This results in the quantum circuit being sub-optimal to the quantum machine at the time of actual execution.

IBM Quantum machines are usually calibrated once a day, likely around 12:00am - 2:00am.
Fig.\ref{Fig:SC21_Cross1} shows that we estimate that over 20\% of our studied quantum jobs were compiled with device information from an older calibration cycle but were executed on the machine after a new calibration.
This results in the compilation being potentially sub-optimal.
Note that these are only coarse estimates based on queuing and execution time stamps.

Fig.\ref{Fig:SC21_Cross2} shows a snippet of a circuit compiled with noise-aware mapping, wherein the noise information of physical qubits is incorporated into the optimal mapping from the circuit's logical qubits to the machine's physical qubits.
Two compilations of the same circuit snippet are shown, from two consecutive calibration cycles.
It is evident that the optimal mapping and circuit structure are different.
Thus, using an older mapping can be detrimental to the fidelity of executed applications.

Note that in the above gate-based compilation approach, the quantum gates are converted to pulses at the time of execution. 
Thus the system will presumably use the most-recently-calibrated pulses to execute the gates on the quantum machine i.e. after the job reaches the head of the queue and is ready for actual quantum execution. 
On the other hand, in the pulse based approach (eg. OpenPulse~\cite{pulse-alexander2020qiskit,pulse-mckay2018qiskit,pulse-gokhale2020optimized}), pulses are generated at the time of compilation.
Thus, these pulses are generated based on machine characteristics at the time of compilation.
A calibration cross-over would mean that even the pulses are sub-optimal at the time of quantum execution.

\subsection{Summary and Recommendations}
    \circled{1}\  \emph{Lack of discipline in load distribution leads to very widely varying queuing times. 
    Queuing time will grow more challenging to predict, as demand, supply and prioritization techniques continue to grow. 
    Research on predicting queuing times with quantitative confidence levels, as pursued in HPC ~\cite{brevik-2006}, are worth pursuing.}
    
    \circled{2}\  \emph{Long queuing times can not only reduce system throughput, but also reduce application fidelity - for example by making device-aware compilations stale. 
    Apart from potentially overlapping long queuing times with nice to have compiler passes, dynamic circuit re-compilation based on new calibration data or monitoring of machine characteristics seems promising. Such instantaneous optimizations would be particularly useful for the pulse-based compilation approach.}
    
    \circled{3}\  \emph{Section \ref{VAF} discussed the use of application metrics such as CX depth and CX error rates to predict application fidelity. 
    Users can be allowed to trade-off fidelity for low queuing time and vice-versa, based on the targets SLAs and the usecase for the quantum application.}
    
    \circled{4}\  \emph{Access privileges, user-defined heuristics, machine popularity and many other characteristics result in unsatisfactory load distribution among machines. 
    Load balancing across machines, especially if performed by the vendor, based on robust machine characterization, can help alleviate large queues, improve system throughput as well as improve user application fidelity.}    
    
    \circled{5}\  \emph{Batching reduces effective per-circuit queuing times. 
    Thus automated techniques at both the client-end and the vendor-end to identify independent circuits, build quantum applications with maximum job parallelism across circuits, seeking inspiration from classical multi-threading techniques~\cite{SMT} etc. will have benefits.}
    
    

%% file: 6_obsD.tex

\section{Trends in Execution}
In this section, we examine execution time across different machines and batch size and explore the ability to predict execution times based on job/circuit and machine characteristics.

\subsection{Execution Time Distribution vs. Machine}

\begin{figure}[h]
\includegraphics[width=\columnwidth,trim={0cm 0cm 0cm 0cm},clip]{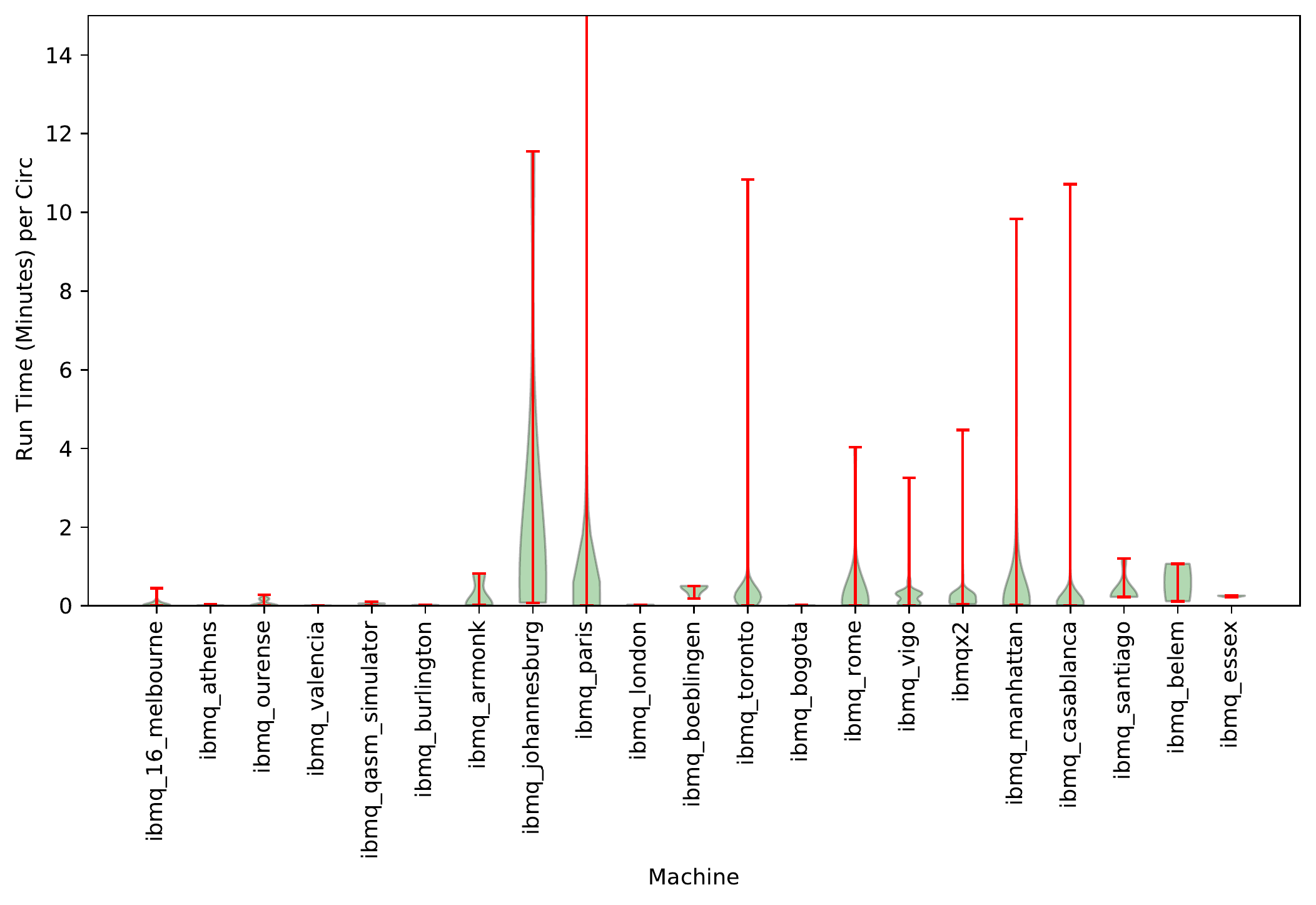}
\centering
\caption{Run time distribution of jobs vs Machine. Run times see non-insignificant variation across circuits} 
\label{Fig:SC21_RT_Machine}
\end{figure}

Fig.\ref{Fig:SC21_RT_Machine} show the distribution of execution time of circuits run on different IBM Quantum Machines.
While run times are significantly lower than queuing times (as discussed prior), the run times still see non-insignificant variation across different circuits - varying from less than a minute to greater than 15 minutes.
These run times variations can be attributed to a combination of the characteristics of the circuit running on the machines (such as number of shots, circuit depth/width etc) as well as the characteristics of accessing the machine itself (such as overheads related to the size of the machine, the memory requirements and so on).
More details about this breakdown is discussed in Section \ref{ExePred}.
A common trend that can be observed is that the larger machines have higher run times - as discussed above, this is likely related to both circuit and machine characteristics. Larger machines would mean possibly larger circuits (a common user heuristic) as well as greater overheads.

\subsection{Execution Time Distribution vs Batch Size}

\begin{figure}[t]
\includegraphics[width=\columnwidth,trim={0cm 0cm 0cm 0cm},clip]{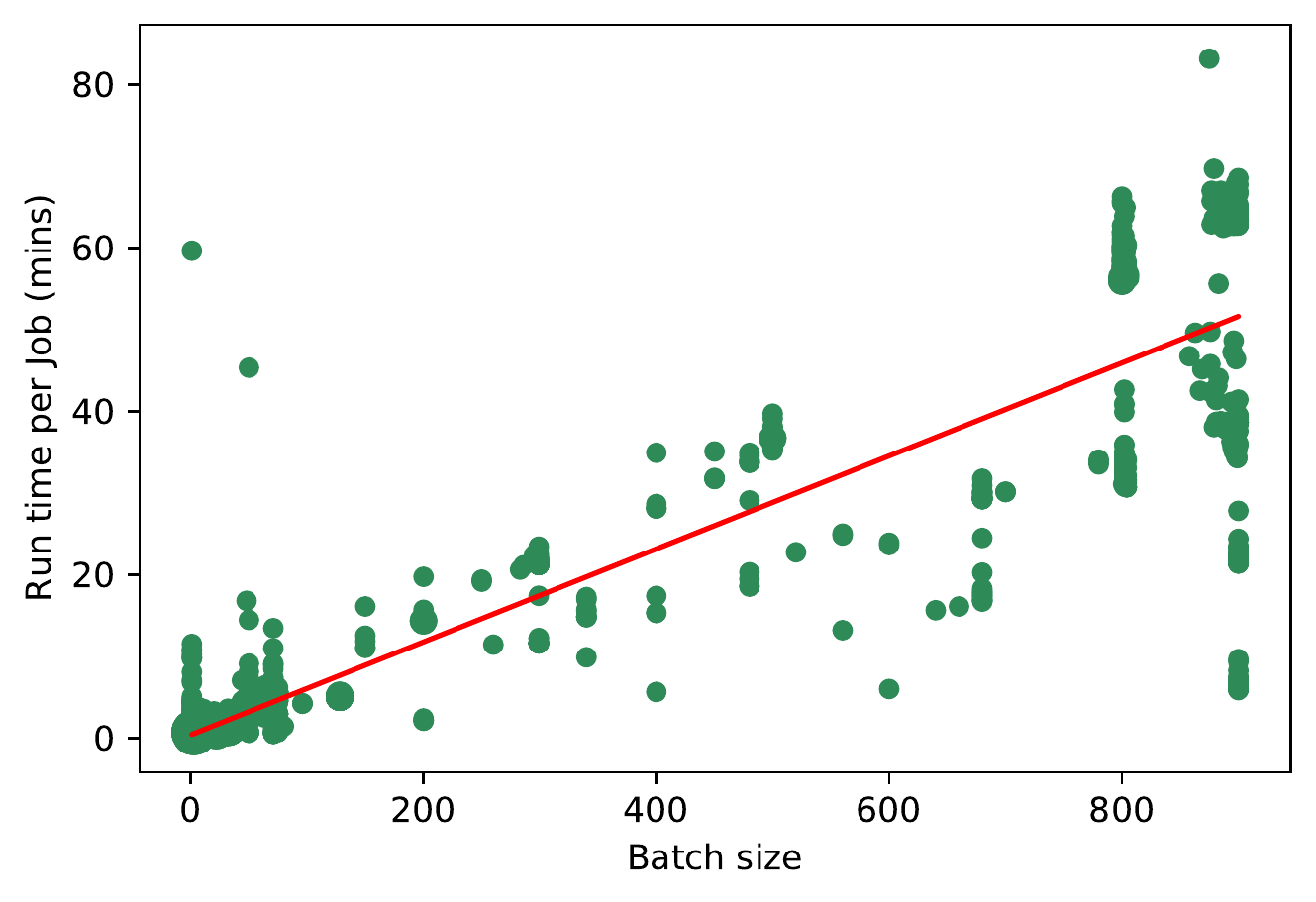}
\centering
\caption{Execution time distribution of jobs vs batch size.  Job runtimes increase proportionally with batch size - the red line.} 
\label{Fig:SC21_RT_Batch}
\end{figure}

Next, Fig.\ref{Fig:SC21_RT_Batch} plots the relation between job runtimes and batch size of the jobs.
Variability in batch size was discussed prior in Section \ref{QTBatch}.
Clearly job runtimes increase proportionally with batch size. 
This is intuitive because more the circuits in the batch, longer is the quantum execution time since the circuits in the batch are executed individually one after the other.
It should be noted that there is considerable overlapping among the green dots closer to the red trend line which are not evident in the figure.
Deviation from the trend (red line) are likely caused by other characteristics such as number of shots and other circuit / machine characteristics and their effects as discussed earlier.
It is possible that independent circuits from one or multiple problems can be executed in conjunction in a batch, enabling lower effective queuing time per circuit.

\begin{figure*}[t]
\includegraphics[width=0.9\textwidth,trim={0cm 0cm 0cm 0cm},clip]{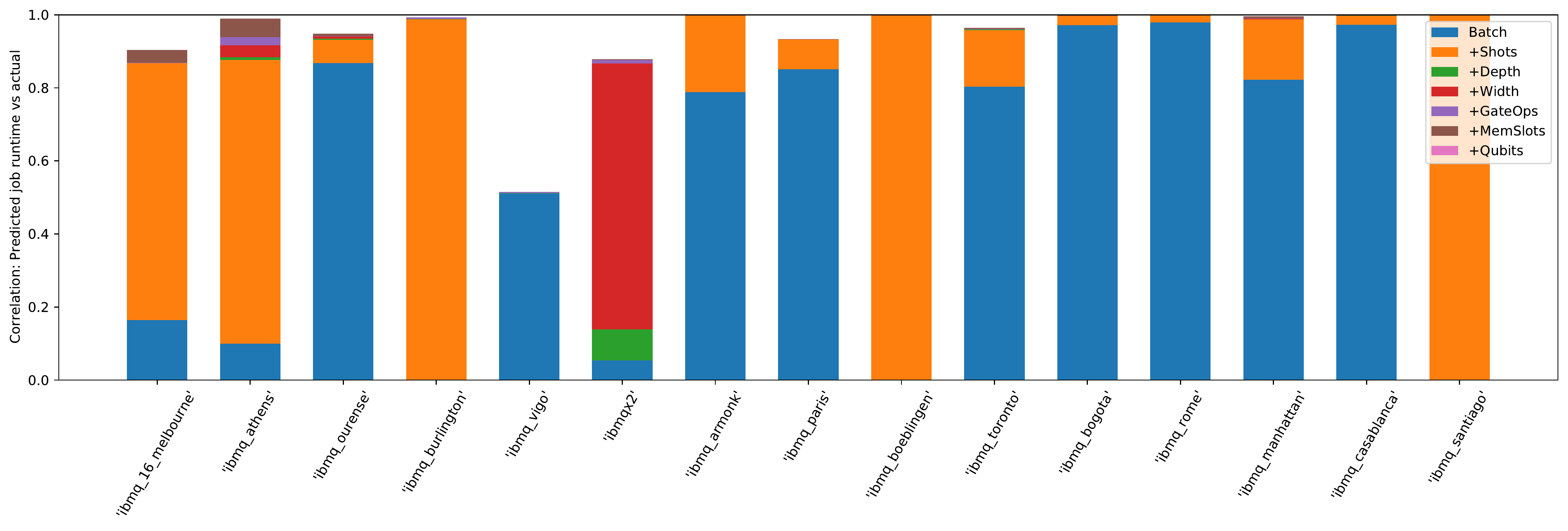}
\centering
\caption{Correlating the predicting runtimes (based on job characteristics) with actual observed runtimes. The major contributor to the correlation is the batch size.
A second contributor is the number of shots.} 
\label{Fig:SC21_RT_Pred}
\end{figure*}

\begin{figure}[t]
     \centering
     \begin{subfigure}[b]{0.48\columnwidth}
         \centering
         \includegraphics[width=\textwidth]{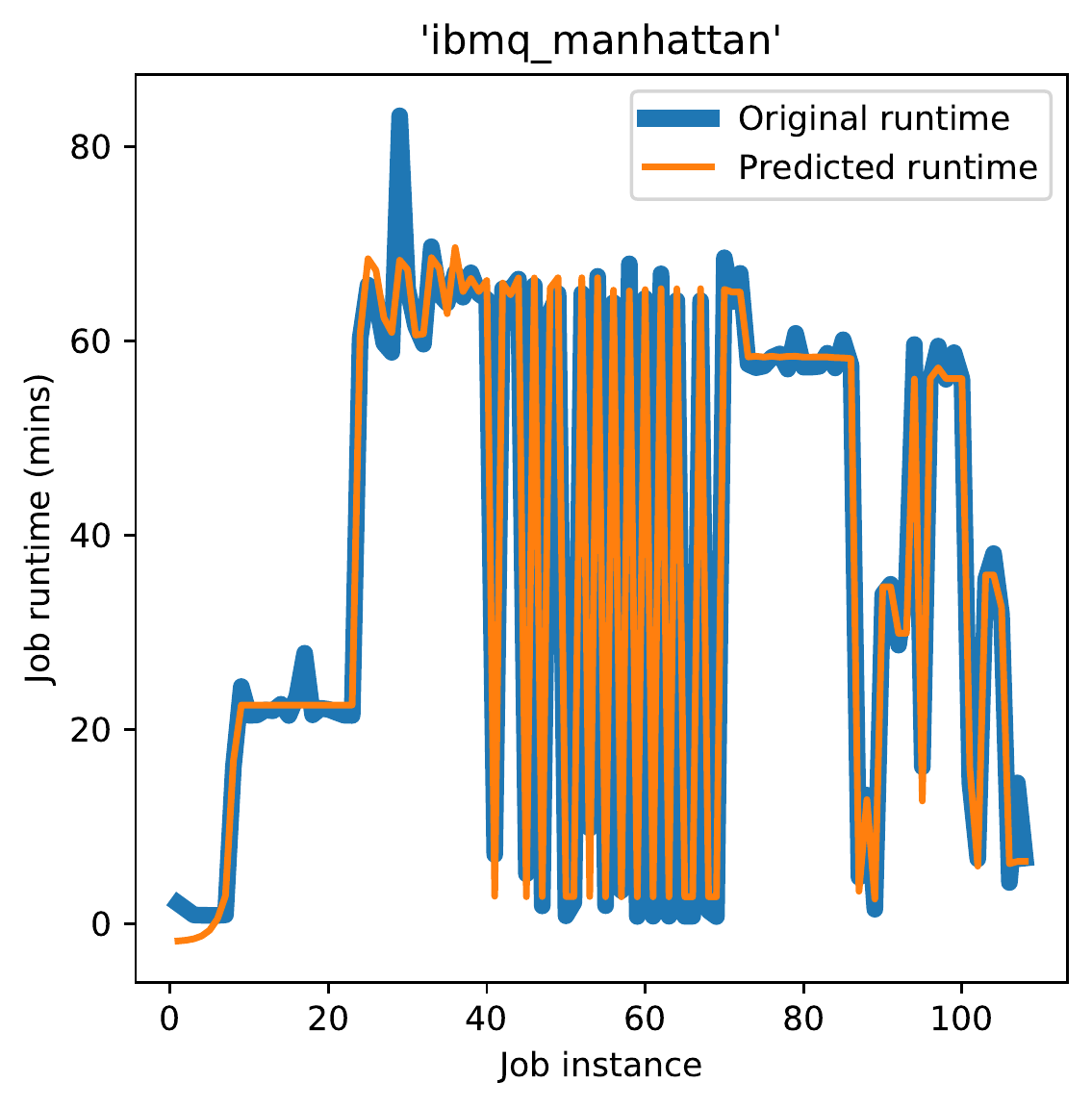}
     \end{subfigure}
     \begin{subfigure}[b]{0.48\columnwidth}
         \centering
         \includegraphics[width=\textwidth]{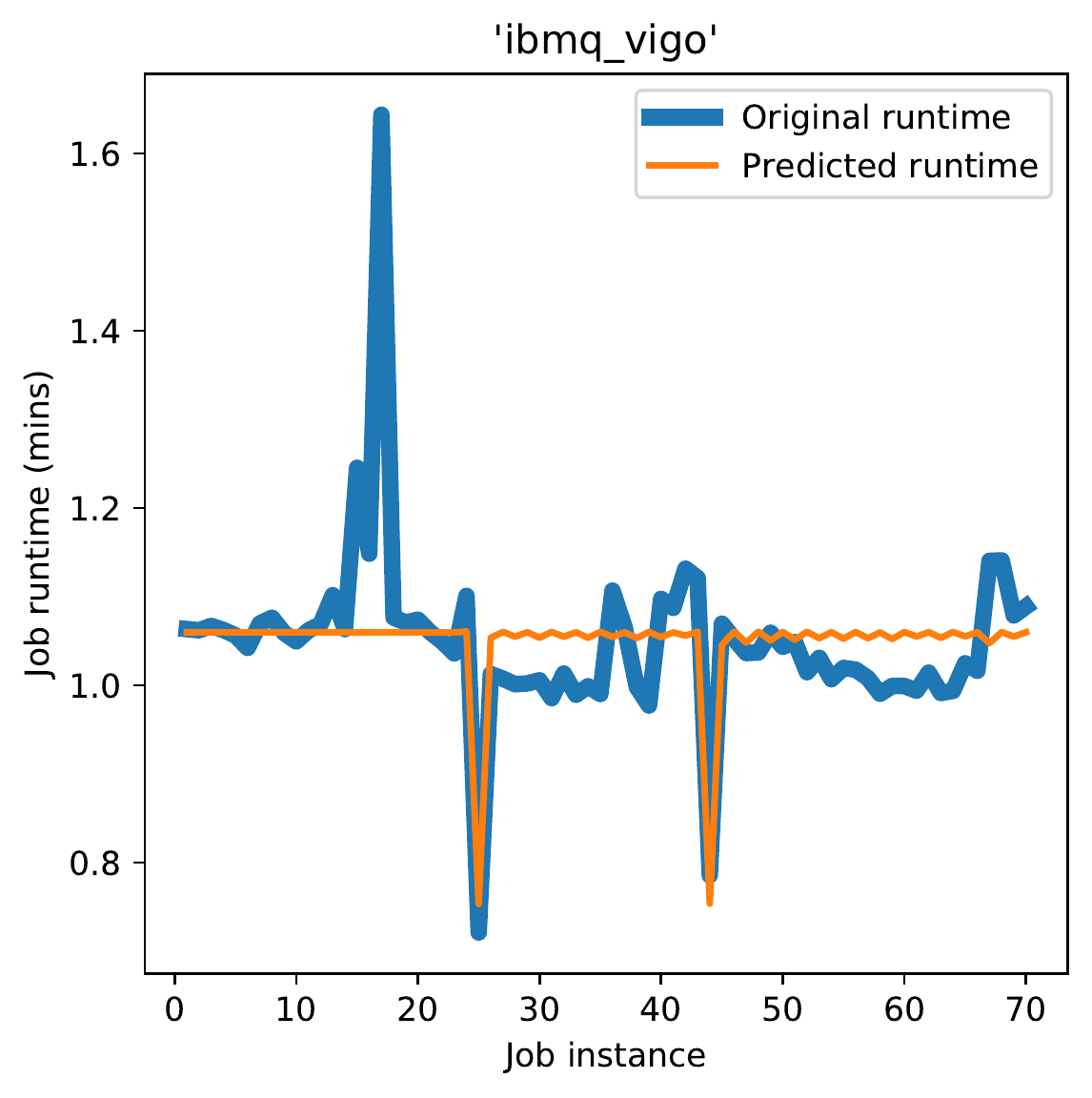}
     \end{subfigure}
        \caption{Predicted / Actual runtimes on different machines. Manhattan shows high correlation while Vigo is poorer.  }
        \label{SC21_PRTred_MV}
\end{figure}

\subsection{Assimilating / Predicting Execution Times}
\label{ExePred}
To understand the dependencies of execution time on characteristics beyond batch size, we build a simple prediction model to predict the execution time of a job based on multiple characteristics.
The model is built as a product of linear terms: $\Pi(a_i + b_i*x_i)$ where $x_i$ is the feature and $a_i$ and $b_i$ are the tuned coefficients.
The studied features are execution: batch size, number of shots; circuit: depth, width  and total quantum gates; and machine overheads: size (proportional to qubits) and memory slots required.
The model is developed with \emph{scipy.optimize curve\_fit}.
Collected data is split into training and test sets (70 / 30\%) to build the model.

Fig.\ref{Fig:SC21_RT_Pred} plots the correlation of predicted runtimes vs actual runtimes, averaged across all jobs that ran on each quantum machine.
Correlation is calculated with the Pearson Coefficient.
First, note that the correlation is 0.95 or above on all but two machines.
The major contributor to the correlation is the batch size, which follows from the analysis in the previous section.
A second contributor is the number of shots which is usually influential when the batch size of the job is low.
Runtimes increase with shots, but at a fractional rate.
Other factors like depth, width and memory slots have limited influence - mostly suggesting that batching and shots are the main contributors in the common case.

In Fig.\ref{SC21_PRTred_MV} we plot the actual runtimes for different jobs on a particular machine in comparison to the predicted runtimes.
For IBMQ Manhattan, the high accuracy in prediction is evident.
We also plot IBMQ Vigo which showed the lowest correlation.
The low correlation is primarily because the range in job runtimes are fairly low, and the correlation is relative to this runtime range.
Thus a 1-minute prediction error is reflected poorly in the correlation.  

Thus, while machine and job characteristics can vary widely, application's runtimes remain fairly predictable.
This is primarily because we are in the early stage of quantum computing exploration in which the number of qubits are low and the algorithmic depth and complexity of the circuits are limited.
Therefore the overheads associated with execution of a circuit is more influential than the characteristics of the circuit itself - this trend is expected to persist in the near term.
The predictability of execution time means that means that building schedulers optimized to different constraints are feasible in the near-term with very low overheads.

In comparison, execution performance prediction is more challenging in classical applications.
At a high level, performance speedups can be determined via Amdahl's law~\cite{amdahl1967validity}, extensions to the multi-core era~\cite{hill2008amdahl} and roofline modeling for heterogeneous systems~\cite{gables}.

Also note that though quantum circuit characteristics like width and depth do not significantly affect execution time, they considerably affect compilation times as observed in Fig.\ref{Fig:PassTime-QFT}.

\subsection{Summary and Recommendations}

    \circled{1}\  \emph{Execution times are considerably lower than queuing times, even though there is variation across jobs and circuits.
    These variations are mostly influenced by machine characteristics rather than job characteristics.
    This is because the current complexities of NISQ-era quantum circuits are low enough that machine executions overheads are greater than the actual execution time of the circuit.}
    
    \circled{2}\  \emph{In the near future, execution times are likely to be highly predictable and are mostly dependent on a few characteristics. The ability to predict execution time accurately amplifies the possibility of implementing the prescribed recommendations related to scheduling, predicting queuing times and leveraging queue times for other tasks.}
    
    \circled{3}\  \emph{As complexity of circuits increases, machine-aware execution time predictions inspired by performance estimations techniques in the classical world are worthy of exploration.}
    
    \circled{4}\  \emph{Perceived execution times can be reduced with novel circuit features like mid-circuit measurement~\cite{Honeywell,IBM-Mid}. 
    Such features will add an additional level of complexity to execution time estimations.}

%% file: 9_backend.tex
\section{Discussion}
\textbf{Related Work:}
To our knowledge, this is the first work studying the effects of quantum job and machine characteristics in the cloud.
In the classical domain, multiple works have studied various system and job characteristics of HPC systems~\cite{schla:2016,Rodrigo:2016,osti_2018,Attig:2011,Patel:2020}.
~\cite{Patel:2020,Rodrigo:2016} have analyzed resource allocation of jobs and their impact on wait times - our work performs such analysis for quantum jobs and circuits.

\textbf{Comparisons to classical HPC:}
We discuss recommendations that are unique to quantum computing or have features that have to be uniquely analyzed with QCs in mind:

\circled{1}\ \emph{Incorrect executions:} While verification and incorrect executions are common in the HPC realm, the challenge in quantum computing is that verification techniques are still at a very nascent stage.
Classical techniques for verification of quantum computations can grow in complexity as much as the quantum computation itself and quickly become unverifiable.

\circled{2}\ \emph{Compilation times:} Unlike classical compilation, quantum compilation is deeply tied to the characteristics of the hardware - error rates, topology and its calibrated state. Thus compiled executable cannot be distributed independent of the machine. 
Thus, the penalties of growing compilation times are likely of greater significance than they are in the classical computing domain. 

\circled{3}\ \emph{Fidelity:} Choosing the right machine to maximize the fidelity of quantum computations is unique to quantum computing, and especially complex give the unique interactions of quantum computations with the environment it executes in.

\circled{4}\ \emph{Queuing times:} While reducing queueing times is well researched in the HPC domain, they are even more critical in the quantum domain because of the irregular drift in quantum characteristics as well as regular recalibration of quantum machines. Earlier execution generally has higher fidelity potential.

\circled{5}\ \emph{Execution times:} Execution times in quantum are highly predicable compared to HPC. This is because overheads dominate quantum execution and execution times are primarily proportional to batch size and to some extent the number of shots executed. Higher predictability can allow for better scheduling policies.

\textbf{Scope of this study:}
Our study has focused on data collected across IBM's fleet of quantum computers, from over a two year period in an academic setting.
While the execution data is limited to those circuits run in the study, the queuing data are generally applicable to all users of the IBM quantum systems over the studied period.
Further, as discussed in earlier sections, the execution data is less closely tied to the specifics of the quantum circuits being run and is more tied to the size of the jobs, number of shots etc - all of which again are applicable to all users of these systems.
Similarly, the general impact of calibration, noise characteristics, constraints of device connectivity etc on application fidelity, are not restricted in any way to the specific circuits executed in this study.

Our study has focused on IBM quantum devices, but the findings are generally applicable to other providers.
Increasing demand for machines is not limited to IBM machines, especially thanks to the easy access to different machines through the cloud~\cite{Web-1,Web-2}.
IBM quantum machines are superconducting devices, as are machines from Rigetti, Google etc. 
These machines, in general, are limited in connectivity, more noisy and require frequent recalibration - hence all our results / recommendations should see considerable similarities across these machines.
Trapped-ion based machines from Honeywell and IonQ achieve better connectivity and lower error rates.
They still require calibration and would benefit from appropriate machine choices etc.
Hence our insights on verification, compilation times, managing fidelity, predicting execution times and managing queuing times are broadly applicable, though further studies are recommended.

\section{Conclusion}
In the current NISQ era, there is severe scarcity in the availability of quantum resources with ever growing demands. 
Similar to classical HPC, vendors should try to allocate machine resources as efficiently as possible so as to improve system throughput, while clients should try to make efficient use of job deployment strategies to maximize their allocated time and resources.
To obtain and understand such insights, it is critical to understand the characteristics of the executing quantum jobs as well as those of the machines in the cloud.
In this study, we analyzed quantum executions on more than 20 IBM Quantum Computers~\cite{IBMQE}, in an academic research setting, over a 2 year period up to April 2021.
Based on our insights, we made recommendations and contributions to improve the management of resources and jobs on future quantum cloud systems.